\documentclass[aps,prb,amsmath,twocolumn,amssymb,titlepage,superscriptaddress]{revtex4-2}

\usepackage{siunitx}

\usepackage{textgreek}  % for greek letters in paper titles \textgamma

\usepackage{graphicx}% Include figure files
\usepackage{dcolumn}% Align table columns on decimal point
\usepackage{bm}% bold math
\usepackage[utf8]{inputenc}
\usepackage[T1]{fontenc}
\usepackage{mathptmx}

\usepackage{amsthm, amssymb, amsfonts,amsmath} %amsmath

\usepackage{multirow}

\usepackage{xcolor}
\definecolor{ronredhighlight}{RGB}{201,98,104}

\usepackage{booktabs}
\usepackage{xcolor}

\newcommand\TAL{TA$_\textup{L}$}
\newcommand\TAX{TA$_\textup{X}$}
\newcommand\LAL{LA$_\textup{L}$}
\newcommand\LAX{LA$_\textup{X}$}
\newcommand\WNU{\si{\per\cm}}

\usepackage[normalem]{ulem} % normalen caises that \em and \emph is not redefined, which is used by cite with biblatex (et al.)

\begin{document}

%\preprint{AIP/123-QED}

\title[]{Determination of acoustic phonon anharmonicities via second-order Raman scattering in CuI}
% Force line breaks with \\

%\author{R. Hildebrandt $^1$, M. Seifert$^2$, Janine George$^{2,3}$, S. Blaurock$^4$, S. Botti$^{2,5}$, H. Krautscheid$^4$, M. Grundmann$^1$, C. Sturm$^1$}
%
%\address{1 Universität Leipzig, Felix Bloch Institute for Solid State Physics, Semiconductor Physics Group, Linn\'{e}stra{\ss}e 5, 04103 Leipzig }
%\address{2 Friedrich-Schiller-Universität Jena, Institute of Condensed Matter Theory and Optics, Max-Wien-Platz 1, 07743 Jena}%
%\address{3	Federal Institute for Materials Research and Testing 	Department Materials Chemistry, Unter den Eichen 87, 12205 Berlin}%
%\address{4	Univerität Leipzig, Institute for Inorganic Chemistry, Johannisallee 29, 04103 Leipzig}
%\address{5	Ruhr University Bochum, Research Center Future Energy Materials and Systems of the Research Alliance Ruhr, Faculty of Physics and ICAM, Universitätstra{\ss}e 150, 44780 Bochum}%
%%

\author{R. Hildebrandt}
%\email{ron.hildebrandt@uni-leipzig.de}
\altaffiliation[Contact: ]{ron.hildebrandt@uni-leipzig.de}%Lines break automatically or can be forced with \\
\address{ 
	Universität Leipzig, Felix Bloch Institute for Solid State Physics, Semiconductor Physics Group, Linn\'{e}stra{\ss}e 5, 04103 Leipzig
}

\author{M. Seifert}
\address{%
	Friedrich-Schiller-Universität Jena, Institute of Condensed Matter Theory and Optics, Max-Wien-Platz 1, 07743 Jena
}%

\author{J. George}%Janine George
\address{%
	Federal Institute for Materials Research and Testing
	Department Materials Chemistry, Unter den Eichen 87, 12205 Berlin}%
\address{%
	Friedrich-Schiller-Universität Jena, Institute of Condensed Matter Theory and Optics, Max-Wien-Platz 1, 07743 Jena
}%

\author{S. Blaurock}
\address{%
	Univerität Leipzig, Institute for Inorganic Chemistry, Johannisallee 29, 04103 Leipzig
}

\author{S. Botti}
\address{%
	Friedrich-Schiller-Universität Jena, Institute of Condensed Matter Theory and Optics, Max-Wien-Platz 1, 07743 Jena
}%
\address{%
	Ruhr University Bochum, Research Center Future Energy Materials and Systems of the Research Alliance Ruhr, Faculty of Physics and ICAM, Universitätstra{\ss}e 150, 44780 Bochum
}%

\author{H. Krautscheid}
\address{%
	Univerität Leipzig, Institute for Inorganic Chemistry, Johannisallee 29, 04103 Leipzig
}%

\author{M. Grundmann}
\address{ 
	Universität Leipzig, Felix Bloch Institute for Solid State Physics, Semiconductor Physics Group, Linn\'{e}stra{\ss}e 5, 04103 Leipzig
}

\author{C. Sturm}%
\address{ 
	Universität Leipzig, Felix Bloch Institute for Solid State Physics, Semiconductor Physics Group, Linn\'{e}stra{\ss}e 5, 04103 Leipzig
}

\date{September 15, 2023}% It is always \today, today,
             %  but any date may be explicitly specified

\begin{abstract}

We demonstrate the determination of anharmonic acoustic phonon properties via second-order Raman scattering exemplarily on copper iodide single crystals. The origin of multi-phonon features from the second-order Raman spectra was assigned by the support of the calculated 2-phonon density of states. In this way, the temperature dependence of acoustic phonons was determined down to 10\,K. To determine independently the harmonic contributions of respective acoustic phonons, density functional theory (DFT) in quasi-harmonic approximation was used. Finally, the anharmonic contributions were determined. The results are in agreement with earlier publications and extend CuI's determined acoustic phonon properties to lower temperatures with higher accuracy. This approach demonstrates that it is possible to characterize the acoustic anharmonicities via Raman scattering down to zero-temperature renormalization constants of at least 0.1cm$^{-1}$.
\end{abstract}

\maketitle

\section{Introduction}

Phonons are collective excitation of condensed matter and specifically for solids of fundamental relevance. They are well-defined for crystal lattices, but as well of relevance for non-crystalline solids. The phonon's origin is bond related and is hence of local nature~\cite{Pages.2006}. This may reveal characteristics, which are otherwise concealed for delocalized properties~\cite{Pages.2006}. Interaction of phonons with other particles are of relevance in low-dimensional materials~\cite{Bai.2022}, for electron-phonon coupling in general~\cite{Bai.2022}, in phonon reabsorption processes~\cite{Aksamija.2009} as well as for non-radiative recombination processes~\cite{Kirchartz.2018}. Aside of sound waves, also the transport of heat is mediated via phonons. Here, low thermal conductivity may be used for thermal barriers~\cite{Knoop.2023} or as well in in thermoelectric applications~\cite{Coroa.2019,Schmidl.2022}. These thermal transport properties though, are directly linked to phonon anharmonicities. For a deep understanding of these processes a precise characterization of general phonon properties is required.\par
By using optical spectroscopy phonon properties may be investigated. Observed energetic shifts and broadening of phonon modes in dependence on temperature contain information regarding their respective harmonic and anharmonic properties. This covers for example the lattice expansion and phonon decay processes. Using Raman spectroscopy, this type of analysis is a well known approach for optical zone-center phonons of various materials. Extending this to acoustic phonons is of interest to assign all related processes~\cite{Wei.2021}.\par 	
By this extension, the need of large neutron scattering facilities to characterize zone-edge phonons would be eliminated. This is demonstrated here by using second-order Raman scattering to characterize acoustic phonons regarding their anharmononicities.\par
We will show this exemplarily on copper iodide (CuI), which has at room temperature (RT) a zinc blende crystal structure. It is known for it's strong anharmonic properties~\cite{He.2019,Serrano.2002,Knoop.2023}, which reflects in CuI's coefficient of thermal conduction, which is as low as \SI{0.5}{W m\tothe{-1}K\tothe{-1}} and hence the dimensionless figure of merit $ZT$ is about $0.25$~\cite{Yang.2017,Coroa.2019,Schmidl.2022}. These are promising values for device applications, especially for transparent electronics due to its wide bandgap of 3.1\,eV~\cite{Kruger.2021} and a hole mobility of up to 43\,cm$^2$/Vs~\cite{Chen.2010}. Due to its intrinsic p-type charge carriers, this semiconductor may as well be a suitable complement to the nowadays established wide-bandgap semiconductors ZnO, GaN or SiC, which can be found in high-power or high-frequency applications. CuI has a large exciton binding energy of about 62\,meV which makes this promising for opto-electronic applications~\cite{Kruger.2021}. Device applications in hetero pn-junctions were demonstrated for solar cells~\cite{Lin.2018}, transistors~\cite{Liu.2020} and light emitting diodes~\cite{Baek.2020}. The performance of these different types of electronic and opto-electronic devices is coupled to phonon interactions and especially acoustic phonon properties of interest for thermal or thermoelectric applications.\par
Information regarding CuI's acoustic phonons is up to now only available via neutron scattering at 90\,K and RT~\cite{Hennion.1972} and as best of knowledge of the authors, there exist only one short report of the second-order Raman spectrum up to now~\cite{Prevot.1974}. The accuracy of 5\,\WNU{} is comparable to reported shifts of phonon energies with temperature. Also a systematic deviation of respective determined acoustic phonons can not be excluded, as found for the optical zone-center phonons when compared to Raman scattering results~\cite{Prevot.1974, Brafman.1976}. These uncertainties are significant and not addressed despite the recent research interest in this material.\par 
Here we report on the fundamental acoustic phonon properties, which are determined with high accuracy and resolution at high symmetry points. We investigate the temperature behavior and discuss harmonic and anharmonic contributions to the phonon properties. We achieve a energetic accuracy down to 0.3\,\WNU{} and temperatures as low as 10\,K, which is a significant improvement compared to available data. The paper is structured as follows: Relevant experimental and computational methods are described in Sec.~\ref{sec:Methods}. 
In Sec.~\ref{sec:Results_and_discussion} we discuss the phonon assignment and energy shift of the second-order modes, which are then compared to the results of the DFT calculations. Finally, this is used to analyze the harmonic and anharmonic contributions of acoustic phonons. The results are summarized in Sec.~\ref{sec:conclusion}.

\section{Methods}
\label{sec:Methods}

\subsection{Crystal growth and quality}
By using copper acetate monophydrate (Sigma-Aldrich, 98\%), acetone and iodine (Sigma-Aldrich, 99.5\%), copper iodide was synthesized in hot acetic acid. The procedure was similar to the one reported in Ref.~\cite{Hardt.1965}. The purification of the crude product, was carried out by crystallization, followed by thermal decomposition of an acetonitrile copper iodide complex [(CH$_3$CN)$_2$(CuI)$_2$]$_\textup{n}$~\cite{Kruger.2021}. The crystal growth of copper iodide single crystals was realized in autoclaves by taking advantage of the inverse temperature-dependent solubility of copper iodide in acetonitrile~\cite{Gao.2013}, similar to the procedure Described in Ref.~\cite{Kruger.2021}. The resulting $\gamma$-CuI bulk crystals were colorless with typical dimensions in the range of 1-4\,mm.\par
The purity of the crystals was determined to be larger than 99.999\% with a Bruker S2 Picofox TXRF using a multi-element standard.\par
Structural characterization of the crystals was done by \mbox{X-ray} diffraction using a PANalytical X'Pert Pro diffractometer with a Cu K$_{\alpha}$. The crystal facets were (111)-oriented and a cubic lattice constant of $a=$6.053(3)\,\r{A} was determined. $\phi$-scans of asymmetrical reflections indicated no twinning or rotational domains of the crystals. Details of the procedures and \mbox{X-ray} diffraction scans can be found in Ref.~\cite{Kruger.2021}.

\subsection{Raman scattering}
All Raman measurements were performed in a helium flow cryostat (Janis ST-500) with the CuI single crystal fixed on a c-plane sapphire substrate. The temperature was determined by using a calibrated silicon diode mounted on the cold finger. As excitation source we used a 100\,mW diode pumped solid-state laser ($\lambda=532.06\,$nm, Coherent Compass 315M). A Mitutoyo microscope objective with an $NA\,=\,0.42$ was used in a back scattering configuration and a Glan-Thompson prism including a half-wave plate was used to analyze the linear polarization. A double monochromator (Jobin Yvon U1000) with a focal length of 2\,$\times$\,1m and a 2400 lines/mm grating resulted in a spectral resolution of about 0.2\,\WNU{} (CCD: Symphony II BIUV). One measurement window covers a spectral range of about 115\,\WNU{}, with about 17\% spectral overlap of two distinct measurements. Rotational Raman lines of air were observed (removed later from spectra) and used for calibration of Raman shift features up to 150\,\WNU{}.

\subsection{Computational details}
The computations were done in the framework of density functional theory (DFT) using the Vienna {\it ab initio} simulation package VASP~\cite{Kresse.1996, Kresse.1999} with the projector-augmented wave method~\cite{Bloechl.1994}. The $4s$ and $3p$ electrons of Cu and $5s$ and $5p$ electrons of I are thereby treated explicitely as valence electrons. Starting with the pure zincblende $\gamma$-phase, we used a cutoff energy of 770~eV for the plane-wave basis set and $\mathbf{k}$-point grid of 8$\times$8$\times$8. We used here tighter computational settings than in our other studies on CuI~\cite{Seifert.2021, Seifert.2022}, to avoid spurious imaginary modes around $\Gamma$-point as a potential result of numerical noise~\cite{Pallikara.2022}. We have applied the exchange-correlation functional PBEsol~\cite{Perdew.2008} for structural optimizations and phonon computations. For the full structural optimization with DFT at 0~K, we relaxed all forces until they were smaller than 10$^{-5}$~eV/\r{A}. This led to a lattice parameter of 5.940~\r{A} for PBEsol which is quite close to the experimental value of 6.053~\r{A}. For the calculation of the Born charges for the non-analytical corrections (NAC), we used a denser $\mathbf{k}$-point grid of 16$\times$16$\times$16.\par
For the phonon calculations, we used the package phonopy~\cite{Togo.2008, Togo.2015} together with VASP and computed the harmonic phonons with the help of the the finite displacement method with a displacement of 0.01~\r{A}. We used 4$\times$4$\times$4 supercells of the primitive cell with a $\mathbf{k}$-point grid of 2$\times$2$\times$2 for the computation of the forces. Information on the convergence of the supercell size are given in the supplementary. For the calculation of the 1-Phonon density of states (1PDOS) and the thermal properties, we used a $\mathbf{q}$-point mesh of 16$\times$16$\times$16. For the broadening of the 1PDOS, we used a gaussian smearing with a full width half maximum (FWHM) of 2~cm$^{-1}$. To compute the temperature-dependence of the phonon modes, we relied on the quasi-harmonic approximation~\cite{Stoffel.2010}. For this we chose different volumes roughly around $\pm$ 10\% of the DFT minimum volume. All these cells were relaxed with the constraint to keep the volume constant. We then performed phonon calculations with consistent settings, mentioned above, for these different volumes around the DFT minimum volume to obtain a description of the free energy as a function of volume at different temperatures. Based on this, we obtain the temperature-dependent equilibrium volumes and repeated the phonon calculations at those volumes corresponding to certain temperatures to compare the calculated results with the experimental derived ones. Further, the obtained results of the phonon properties were compared with the literature to benchmark the chosen parameters and therefore validate these results for the calculation of the 2-phonon density of states. More details regarding this procedure are shown in the supplementary. \par
The 2-phonon density of states (2PDOS) was obtained out of the phonon frequencies following the equation as defined by Ref.~\cite{Okubo.1983}:
\begin{eqnarray}
	\textup{2PDOS}(\vec{q},j) &=& 2\pi \sum_{\vec{q}',\vec{q}''} \sum_{j', j''} \Delta\left( \vec{q} - \vec{q}' -\vec{q}'' \right)	\times \\
	&&\delta \left( \omega \left(\vec{q},j\right) - \omega\left(\vec{q}',j'\right) - \omega\left(\vec{q}'',j''\right) \right) \nonumber \label{eq:2PDOS}
\end{eqnarray}
$\Delta(\vec{Q})$ is 1, if $\vec{Q}$ is either zero or a multiple of the reciprocal lattice vector.  Otherwise $\Delta(\vec{Q})$ is zero. Furthermore $\vec{q}$ denotes the phonon wave vector, $j$ the corresponding phonon mode and $\omega$ the phonon frequency. After evaluating Eq~(1), just those $\vec{q}$-points where considered which are located in a small environement around the $\Gamma$-point to account for the Raman scattering which is limited to $\Gamma$-point. Therefore only $\vec{q}$-points inside a box of length $2/8|q_{max}|$ and the Gamma-point in its centre were considered. The 2PDOS was broaded with a Gaussian smearing, with a full width half maximum (FWHM) of 2~cm$^{-1}$.

\section{Results and Discussion}
\label{sec:Results_and_discussion}
\subsection{Second-order Raman features}
\label{sec:Second_order_Raman_features}
\begin{figure*}  \centering
	% in C:\Daten\2 My contributions\Paper\2022 1 Second Order Raman CuI\Plots_for_CuI_Second_Order_Raman_Paper.opj --> 11K Spectrum Publication
	\includegraphics[width=0.980\textwidth]{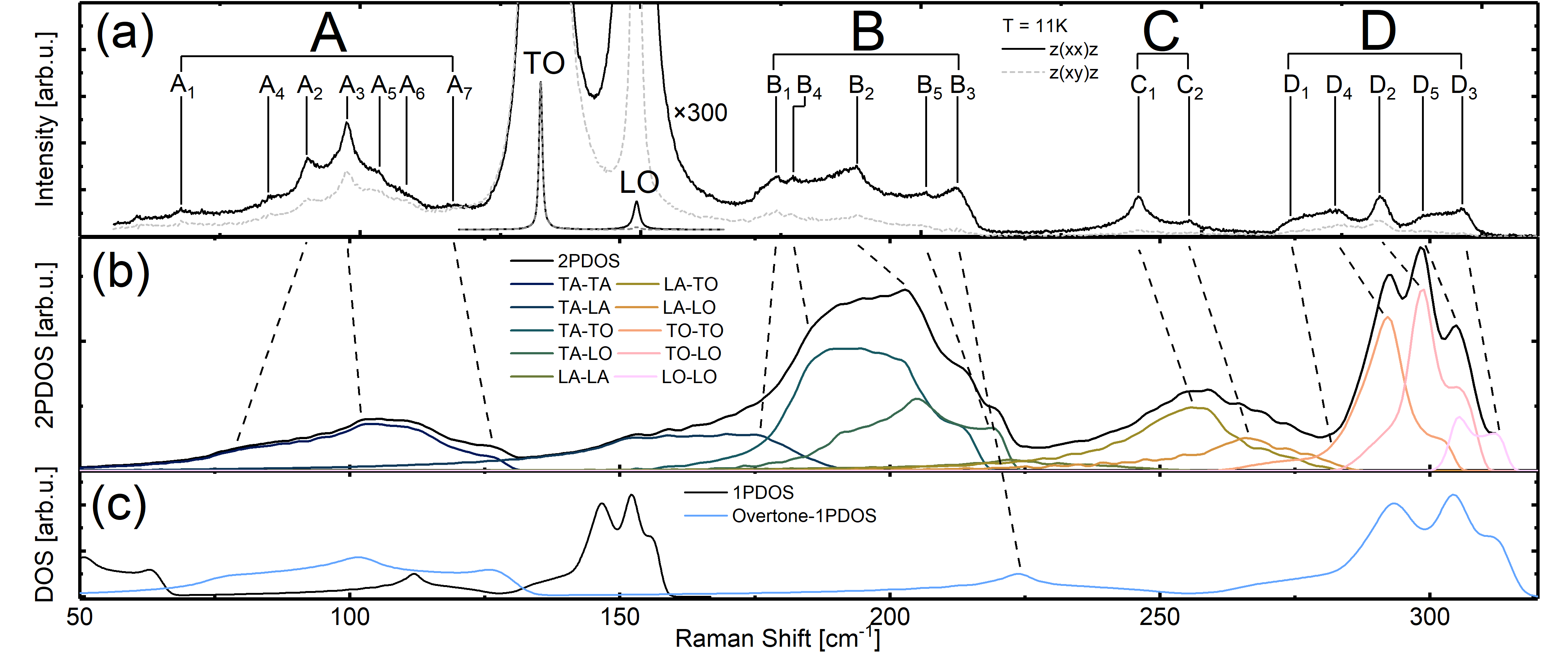}
	\caption{\label{fig:raman_spectrum_9K_with_calc} (a) Second-order Raman spectrum of CuI at 11\,K for parallel (black, z(xx)\=z) and cross (grey, z(xy)\=z) polarization. The TO and LO mode are as well depicted and rescaled by a factor of 300 (and vertically shifted for clarity). Due to depolarization by the cryostat window, the LO mode does not vanish in cross polarization configuration and is only reduced to 1:13. Four large structures A, B, C and D are observed whereby each fine feature is labeled and listed with details in Tab.~\ref{tab:feature_energies}. (b) 2-phonon DOS calculated via the PBEsol phonon dispersion. (c) PBEsol 1-phonon DOS (1PDOS) and Overtone-1PDOS. The higher energy of calculated features compared to the observed experimental ones is related to larger optical phonon energies predicted by PBEsol.}
\end{figure*}
The millimeter-sized single crystals were cooled down to about 10\,K and Raman spectra with different integration time in parallel-\,(z(xx)\=z) and cross-polarization (z(xy)\=z) were measured to observe the first- and second-order Raman scattering features of CuI.\par
The first-order spectra can be seen in Fig.~\ref{fig:raman_spectrum_9K_with_calc}(a) and show the transversal optical (TO) and longitudinal optical (LO) mode at 133.3\,cm$^{-1}$ and 149.4\,cm$^{-1}$ respectively. The second-order Raman signal is shown as well and rescaled by a factor of 300 for clarity. This signal contains four large structures labeled A, B, C and D with various features which are listed in detail in Tab.~\ref{tab:feature_energies}. From the phonon dispersion in Ref.~\cite{Hennion.1972}, each structure's origin can be estimated for CuI.\par
In a second-order Raman process, two phonons are involved in the scattering process. There are three different processes relevant to identify the feature's origin: overtone-, combination- and difference-phonon modes. The overtone process is characterized by a contribution of two phonons from the same phonon branch, with opposite momenta. If their wavenumber is, for example, $\omega_a$, the resulting observed Raman shift is at 2$\omega_a$. The phonon combination\,($+$) or difference\,($-$) processes are characterized by two phonons of different phonon branches whereby both phonons have opposite momenta. If their wavenumbers, for example, are $\omega_a$ and $\omega_b$, the resulting observed Raman shift is at $\omega_a\pm\omega_b$. Momentum conservation is ensured in each processes via opposite momenta of the respective contributing phonons ($q\approx0$). \par
We now proceed with discussing our assignment of the observed structures, based on CuI's known phonon dispersion~\cite{Hennion.1972}. The observed structure between 75\,\WNU{} and 120\,\WNU{}\,(labeled A) were attributed to overtones of transversal acoustic (TA) phonons, whereas the structure between 270\,\WNU{} and 300\,\WNU{} (labeled D) are attributed to combinations and overtones of optical phonons. The intermediate structures labeled B ($170-210$\,\WNU{}) and C ($235-250$\,\WNU{}) can be attributed to combinations and overtones of optical and acoustic phonons. For the structure B, mostly TA and optical phonons are involved, while for the structure C, LA and optical phonons participate. Between each structure, a featureless continuum signal is observed. Each feature is here associated with van Hove singularities of the 2PDOS.\par
The selection rules for second-order Raman processes in zincblende crystals yield that all combinations and overtones are allowed~\cite{Birman.1963} and hence the information of the full phonon dispersion is contained in the second-order Raman spectrum~\cite{Cardona.1981,Loudon.2001,Irwin.1970}.\par
\begin{table}
	\caption{\label{tab:feature_energies} Labels and position of observed Raman features. Assignments of involved phonons are done with the position in the brillouin zone by critical point labels. Along the $\Sigma$-direction, these additional energies were assumed: 
		LA$_{\Sigma}$ = (LA$_\textup{X}$ + LA$_\textup{L})/2=114$cm$^{-1}$, TA$_{\Sigma,\textup{l}}$ = (TA$_\textup{X}$ + TA$_\textup{L})/2=47$cm$^{-1}$, TA$_{\Sigma,\textup{u}} = 61$cm$^{-1}$ and O$_{\Sigma}=$(\,TO$_{\textup{X}}$+\,LO$_{\textup{X}}$)/2. The indices "l" and "u" indicate the lower and upper TA phonon branch along the $\Sigma$ direction. The third column contains the second-order phonon feature energy predicted by our suggested critical point phonon energies from Tab.~\ref{tab:phonon_energies}. All energies are given in cm$^{-1}$.}
		\begin{tabular}{ccccc}
			\toprule
			Label           & Experiment                         &                Prediction via             &      Assignment      &    Point     \\
			&            [cm$^{-1}$]             & Tab.~\ref{tab:phonon_energies} [cm$^{-1}$]&                      &              \\ \midrule
			A$_1$           &       \hphantom{11}67.5$^a$        &   \hphantom{$\approx$\,}67\hphantom{.5}& LA - TA$_\textup{l}$ &   $\Sigma$   \\
			A$_4$           &  \hphantom{1.}84$^a$\hphantom{5}   &   \hphantom{$\approx$\,}81.5& O - TA$_\textup{u}$ &   $\Sigma$   \\
			A$_2$           &          \hphantom{1}90.5          &   \hphantom{$\approx$\,}90\hphantom{.5}    &         2TA          &      L       \\
			A$_3$           &          \hphantom{1}97.5          &   \hphantom{$\approx$\,}98\hphantom{.5}&         2TA          &      X       \\
			A$_5$           &          103\hphantom{.5}          &                     -                  &         2TA          &      K?      \\
			A$_6$           &          108\hphantom{.5}          &                     -                 &         2TA          &      W?      \\
			A$_7$           &    \textgreater116\hphantom{.5}    &   \hphantom{$\approx$\,}122\hphantom{.5}&  2TA$_{\textup{u}}$  &   $\Sigma$   \\ \midrule
			\multirow{ 2}{*}{B$_1$}           &             \multirow{ 2}{*}{173.5$^a$}               &   \hphantom{$\approx$\,}173\hphantom{.5}&       TA + LA        &      X       \\ %\cmidrule{3-5}
			&                                    &   \hphantom{$\approx$\,}175\hphantom{.5}& TA$_\textup{u}$ + LA &   $\Sigma$   \\
			\cmidrule{3-5}
			B$_4$ &          177\hphantom{.5}          &   \hphantom{$\approx$\,}181\hphantom{.5}&       TA + TO        &      L       \\
			\cmidrule{3-5}
			&                               &         $\approx$190\hphantom{.5}&       TA + TO        &      X       \\
			B$_2$                 &      188.5                              &         $\approx$193\hphantom{.5} &       TA + LO        &      X       \\
			&                                    &   \hphantom{$\approx$\,}190\hphantom{.5}&       TA + LO        &      L       \\
			\cmidrule{3-5}
			B$_5$ &               200.5                &         $\approx$ 200\hphantom{.5}& TA$_\textup{u}$ + O  &   $\Sigma$   \\ %\cmidrule{3-5}
			B$_3$           &          206\hphantom{.5}          &   \hphantom{$\approx$\,}206\hphantom{.5}&         2LA          &      L       \\ \midrule
			C$_1$           &               238.5                &   \hphantom{$\approx$\,}238\hphantom{.5}&       LA + TO        &      L       \\ %\cmidrule{3-5}
			%	C$_2$          &               247.5                &         2LA          &      X       &                    248                     \\
			C$_2$           &               247.5                &   \hphantom{$\approx$\,}248\hphantom{.5}&       LA + LO        &      L       \\ \midrule
			D$_1$           &               265.5                &        \hphantom{$\approx$\,}264.4&         2TO          &   $\Gamma$   \\ %                 &                      &        O + LA        &      X       &     $\approx$ 266
			%             &                                    &        O + LA        &      X       &               $\approx$ 266                \\
			\cmidrule{3-5}
			D$_4$ &          274\hphantom{.5}          &   \hphantom{$\approx$\,}270\hphantom{.5}&         2TO          &      L       \\
			\cmidrule{3-5}
			\multirow{ 2}{*}{D$_2$} &          \multirow{ 2}{*}{282\hphantom{.5}}          &              280, 285, 281.5&       TO + LO        & L,X,$\Gamma$ \\
			&                                    &              282\hphantom{.5}&         2TO          &      X       \\
			\cmidrule{3-5}
			D$_5$ &          291\hphantom{.5}          &                  290, 288&         2LO          &     L,X      \\
			D$_3$           &          297\hphantom{.5}          &        \hphantom{$\approx$\,}298.6&         2LO          &   $\Gamma$   \\ \bottomrule
		\end{tabular}\\
		\raggedright
		$^\textit{a}$~Also observed in Ref.~\cite{Prevot.1974}.
\end{table}
To refine the assignment, we will also include estimates of the intensities from the computed phonon dispersion. The intensity dependence of the individual second-order Raman features is quite complex. It depends upon the phonon dispersion, the second-order Raman susceptibilities of the phonon branches as well as their population. Usually, the spectral shape can be approximated by using the overtone phonon density of states (Overtone-1PDOS), which is the phonon density of states (1PDOS) multiplied by a factor of 2~\cite{Cardona.1981}. This especially emphasizes overtone processes, whereby combination processes are only represented in the 2-phonon DOS (2PDOS, calculated via Eq.~(\ref{eq:2PDOS})). The 2PDOS and the Overtone-1PDOS are shown in Fig.~\ref{fig:raman_spectrum_9K_with_calc}(b,\,c). Considering both contributions, the origin of each feature can be assigned and, in this way, the individual phonon energies can be derived.\par 
For all observed features, the involved phonons as well as their energy and the corresponding wave vector in the brillouin zone are summarized in Tab.~\ref{tab:phonon_energies}. For TA phonons, the L and X critical points are easily observable, while for the LA phonons the high density of states favors the observation of \LAL{} related overtones and combinations. Hence, the \LAX{} mode can be neglected in most cases, because of its low density of states. This is similar to ZnS, which has a quite comparable phonon dispersion compared to CuI~\cite{Serrano.2004,Vagelatos.1974} due to the relative atomic masses and their high transversal charge values~\cite{LaCombe.1971}. The \LAL{} DOS dominance is as well indicated by the phonon dispersion in Ref.~\cite{Vardeny.1977}. This is also supported by a strong parallel-polarized signal at the B$_3$ feature, which can be a characteristic for overtones~\cite{Cardona.1981}. Due to the low dispersion of the optical phonons, all their combinations and overtones are clearly observed as well. In this way, most features were uniquely identified, while only three features are ambiguous.\par
Deviations from the observed peak position and the phonon energy at the critical point can be seen for example clearly at the D$_3$ feature. This feature D$_3$ (297\,cm$^{-1}$) shows a deviation to its LO$_\Gamma$ overtone assignment (298.6\,cm$^{-1}$) by 1.6\,cm$^{-1}$. Deviations of similar magnitude can be observed for other assignments and are usually lower than 2\,cm$^{-1}$. Those deviations can be explained by the different type of maxima involved in the respective 2-phonon scattering transitions and the flatness of the phonon dispersion at respective brillouin zone points. This would reason a deviation of the 2-phonon density of states maximum~\cite{Johnson.1964}, compared to the position from the critical points of the phonon dispersion.\par
From the observed second-order Raman features, the phonon energies at certain critical points in the brillouin zone were deduced and were given in Tab.~\ref{tab:phonon_energies}. Within respective error bars, We observed a good agreement with values determined by neutron scattering~\cite{Hennion.1972,Vardeny.1977}, whereby our values were determined with much higher precision. For acoustic phonons, only the TA phonon at the L-point were observed at higher wavenumber. The energetic ordering of the optical phonons at the respective critical points is resembled in addition to assuming a non-crossing behavior of the TO and LO mode, as observed for the phonon dispersions in Ref.~\cite{Vardeny.1977}. This is expected for CuI due to its high transversal charge similar to ZnS~\cite{Serrano.2002,Plendl.1966,Keyes.1962,Mitra.1963,Vagelatos.1974}. The non-crossing, though, is not reproduced by our calculated phonon dispersion. Similar deviations between experimental and calculated phonon dispersion are as well observed for CuBr or ZnTe~\cite{Petretto.2018,Vagelatos.1974,Vardeny.1977}. Nevertheless, the spectral shape is well resembled by the 2PDOS.\par
For the optical overtones (structure D), the present crossing of TO and LO phonon in the DFT-based phonon dispersion reasons a weak separation of individual TO-TO (D$_1$), TO-LO (D$_2$) and LO-LO (D$_3$) phonon combinations. This is better represented by the individual parts of the 2PDOS (Fig.~\ref{fig:raman_spectrum_9K_with_calc}(b)). \par

%\begin{table}
%	\caption{\label{label}Table caption.}
%	\begin{indented}
%		\item[]\begin{tabular}{@{}llll}
%			\br
%			Head 1&Head 2&Head 3&Head 4\\
%			\mr
%			1.1&1.2&1.3&1.4\\
%			2.1&2.2&2.3&2.4\\
%			\br
%		\end{tabular}
%	\end{indented}
%\end{table}

\begin{table}
	\caption{\label{tab:phonon_energies} Phonon mode energies at critical points $\Gamma$, L and X determined from the second-order Raman scattering spectrum at 9\,K. The last columns are our estimated accuracy of the respective energies considering measurement inaccuracies and deviations resulting from the 2PDOS shape. Note that the results from neutron scattering measurements have a significant larger error bar of about $\pm5$\,cm$^{-1}$~\cite{Hennion.1972}. All values are given in cm$^{-1}$.}
	\begin{tabular}{ccccccccccc}
		\toprule
		Mode & $\omega_\Gamma$  &  $\omega_L$   &  $\omega_X$   &   \hspace{0.5cm} &$\Delta \omega_\Gamma$ & $\Delta \omega_L$ & $\Delta \omega_X$  \\ \midrule
		TA  &   -      & \hphantom{1}45.3  \hphantom{1}\hphantom{1}& \hphantom{1}49.3 \hphantom{1}&                  &       -        &     0.3     &    0.3       \\
		LA  &   -     & 103\hphantom{.1}  \hphantom{1} & 124\hphantom{.1}   &                  &       -        &     2     &    2       \\
		TO  &  132.2\hphantom{1}\hphantom{1}   & 135\hphantom{.1}  \hphantom{1} & 141\hphantom{.1}    &                  &      0.1       &     2     &    4       \\
		LO  &  149.3\hphantom{1}\hphantom{1}   & 145\hphantom{.1}  \hphantom{1} & 144\hphantom{.1}    &                  &      0.1       &     2     &    4       \\ \bottomrule
	\end{tabular}
\end{table}

\subsection{Temperature dependence of Raman modes}
\label{sec:temperature_dependence_of_features}
\begin{figure*} \centering
	\includegraphics[width=0.98\textwidth]{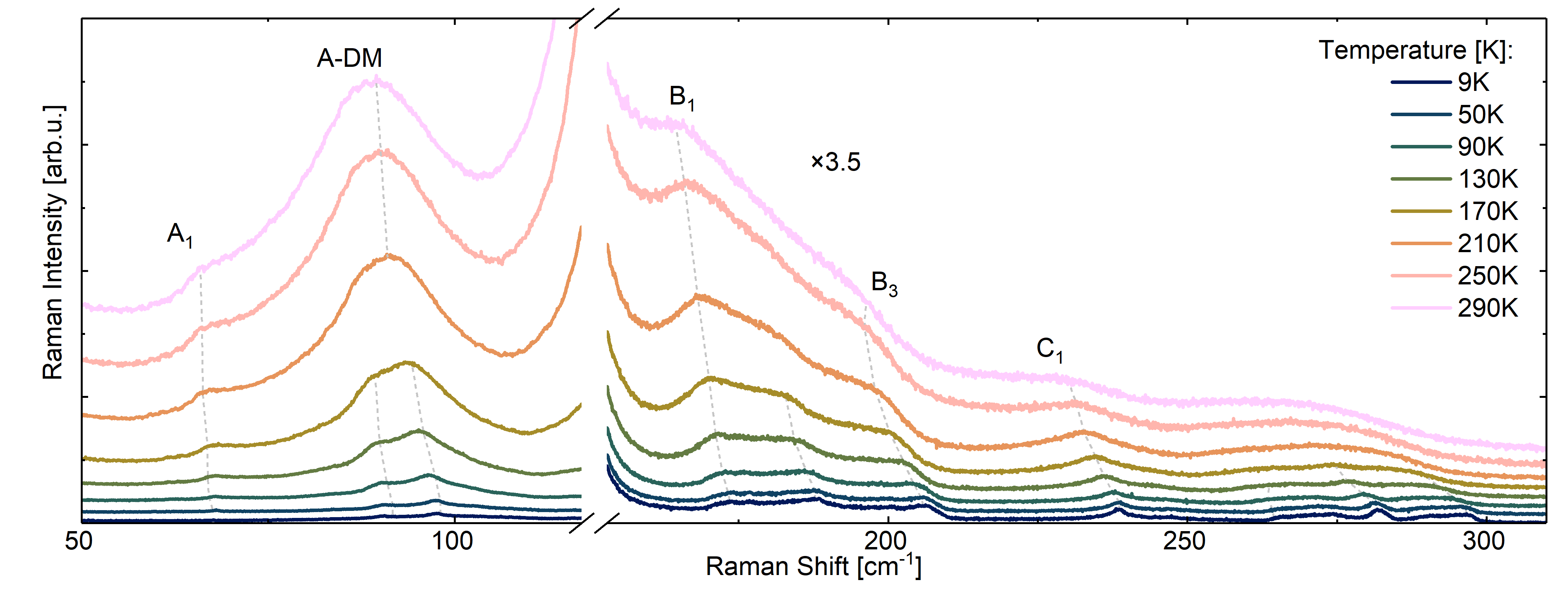}
	\caption{\label{fig:raman_spectrum_vs_temp} Temperature-dependent unpolarized second-order Raman spectra of CuI up to room temperature in 40\,K steps. The intensity of the features above the horizontal break are scaled up by 3.5 for better visibility. Some prominent features at different temperatures are indicated by grey dashed lines and/or labels. The A-DM mode is a difference mode and is not related to the TA overtones. The former is only observable at elevated temperatures, while the latter dominate at 9\,K.}
\end{figure*}

The identified features in the second-order Raman spectrum at low temperature were related to the combinations, differences and overtones of various acoustic and optical phonon modes. With increasing temperature a Raman shift to lower wavenumbers of the features is observed in addition to a strong increase of broadening. This restricts the number of traceable features at elevated temperatures. The measured Raman spectra as a function of temperature are shown in Fig.~\ref{fig:raman_spectrum_vs_temp}.\par
The intensity of the structures B, C and D rises with increasing temperature according to $I_\text{sum} \propto [n(\omega_a)+1][n(\omega_b)+1]$, as expected for overtone and combinations modes~\cite{Potts.1972,Weber.2000}. Here $n(\omega_i)~=~[\textup{exp}(hc\omega_i/k_\textup{B}T)-1]^{-1}$ is the Bose-Einstein distribution function with $\omega_i$ the wavenumber of the phonon. For the structure A at elevated temperatures, an additional contribution from a difference mode has to be considered, i.e. $I_\textup{A} \propto I_\text{sum} +  [n(\omega_c)+1][n(\omega_d)]$ with $\omega_c - \omega_d \approx 90$\,\WNU{}. This difference mode contribution vanishes for low temperatures and can be neglected at 10\,K. Though, at RT this mode is easily observable at about 90\,\WNU{}~\cite{Serrano.2002} and is hence labeled A-DM (difference mode). Its intensity is then comparable to the TO and LO mode and its origin involves various optical and TA phonons.\par
\begin{figure*}  \centering
	\includegraphics[width=0.980\textwidth]{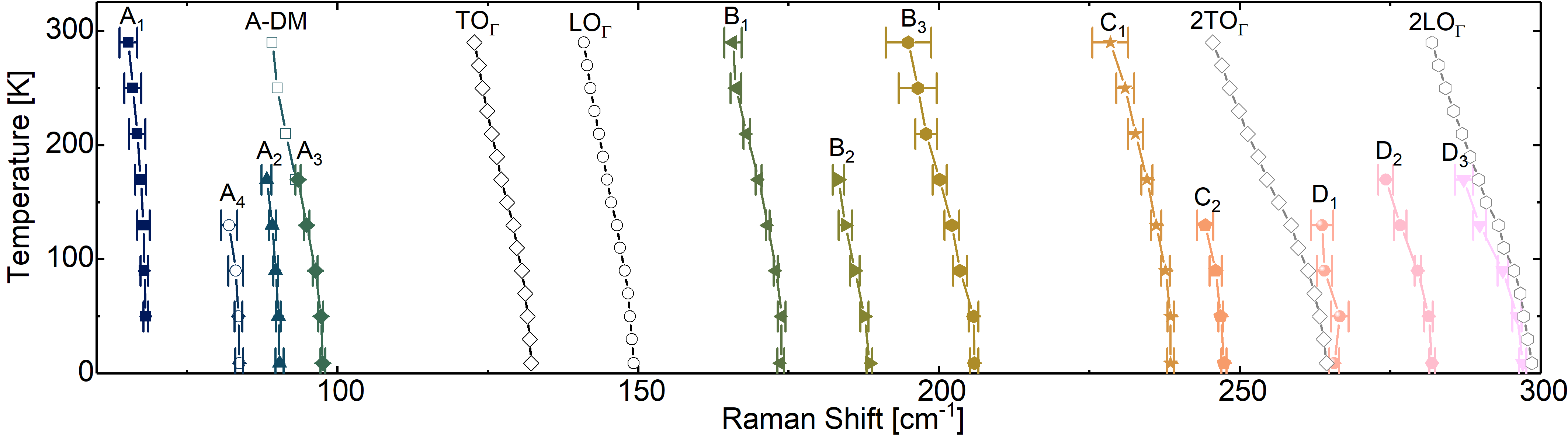}
	\caption{\label{fig:raman_features_vs_temp} Temperature dependence of second-order Raman scattering features and the fundamental TO and LO mode including their expected overtone positions. The A-DM mode (not listed in Tab.~\ref{tab:feature_energies}) is related to various difference processes and has comparable intensity at RT as the fundamental modes.}
\end{figure*}

Due to the strong broadening, most features are only observable up to about 100\,K or 200\,K, respectively. The energy as a function of temperature is shown for all respective features in Fig.~\ref{fig:raman_features_vs_temp}.\par
We assign the features A$_1$ and A$_4$ as difference modes. This assignment was supported by the observed temperature dependencies of their intensities~\cite{Potts.1972,Weber.2000}. The structure D shifts twice as fast as the fundamental TO and LO modes, as expected for overtone and combination modes.\par
With the known origin of the features (Tab.~\ref{tab:feature_energies}) and their temperature dependence, the temperature dependence of acoustic phonons can be determined, which are otherwise not accessible via first-order Raman scattering. The energy of the TA phonons at the L- and X-point is given by their overtones represented by feature A$_2$ and A$_3$ respectively. The LA$_\textup{X}$ phonon can be related to B$_1$ - A$_3$/2 while LA$_\textup{L}$ is related to B$_3$/2.\par
Before the acoustic phonons' temperature dependence is derived, the impact of the lattice expansion and the phonon-phonon interaction on the properties of the optical phonons at the $\Gamma$-point is shortly discussed. This will later support the validity of derived anharmonic contributions, observed in the shifts of acoustic phonons.
\subsection{Optical phonons}
\label{sec:optical_phonons}
The fundamental TO and LO phonons at the $\Gamma$-point were analyzed by means of Lorentzian functions. This data (shown in the supplementary) was described by the Klemens model, in which the Raman shift and the mode broadening are related to harmonic effects of the crystal lattice and anharmonic phonon-phonon interactions, i.e., one phonon decays into multiple other phonons~\cite{Klemens.1966,Giehler.2001,Balkanski.1983}.\par
Considering contributions from the lattice expansion and the decay of phonons into up to three phonons, the Raman shift of the phonon mode as a function of temperature can be described by~\cite{Balkanski.1983,Giehler.2001,He.2019}:
\begin{equation}
	\omega(T) = \omega_0 - \Delta_{\textup{Lat}}(T) + \Delta_{3\textup{ph}}(T) + \Delta_{4\textup{ph}}(T)
	\label{eq:center_fit_eq}
\end{equation}
where $\omega_0$ is the unperturbed phonon energy, $\Delta_{3\textup{ph}}$, $\Delta_{4\textup{ph}}$ are the anharmonic phonon decay processes involving 3 or 4 phonons respectively and $\Delta_{\textup{Lat}}$ is the Raman shift of the phonon due to thermal expansion of the crystal lattice, which is given in detail by~\cite{Giehler.2001,He.2019}:
\begin{equation}
	\Delta_{\textup{Lat}}(T) = 3 \omega_0 \gamma \int_{0}^{T} \alpha(T^{'})\,\textup{d}T^{'}.
	\label{eq:lattice_shift_integral}
\end{equation}
Here $\gamma$ denotes the Grüneisen parameter and $\alpha$ the temperature dependent coefficient of the thermal expansion. The temperature dependence of the thermal expansion coefficient was taken from Ref.~\cite{Nakamura.2022} (see supplementary).\par%Fig.~\ref{fig_sup:alpha_nakamura_and_lattice_constant}
The contributions of the anharmonic processes are described by~\cite{Balkanski.1983,Klemens.1966}:
\begin{eqnarray}
		\Delta_{3\textup{ph}}(T) &=& \ \Delta_3[1+\sum_{i=1}^{2}\nolimits n(\omega_i)] \\
		\Delta_{4\textup{ph}}(T) &=& \ \Delta_4[1+\sum_{j=1}^{3}\nolimits n(\omega_j) + n^2(\omega_j)] 
\end{eqnarray}
with $\Delta_3$ and $\Delta_4$ as zero-temperature renormalization constants describing the strength of the individual contributions to the Raman shift. The Grüneisen parameters $\gamma$ of the respective phonon modes were taken from~Ref.~\cite{Serrano.2002} with $\gamma($TO$_\Gamma) = 1.8$ and $\gamma($LO$_\Gamma) = 2.5$. This is in accordance with other results~\cite{Gopakumar.2017, Plendl.1972}. Our calculated Grüneisen parameter band structure is shown in the supplementary and indicates comparable values. For the description of the mode broadening, a similar model is used. Here the parameters $\Gamma_3$ and $\Gamma_4$ were used, which describe the 3- and 4-phonon processes, respectively~\cite{Giehler.2001,Klemens.1966} (see supplementary).
\begin{table}
	\caption{\label{tab:phonon_decay_fit_params_O} Fit parameters obtained via Eq.~(\ref{eq:center_fit_eq}) and S1 (the fits are shown in the supplementary). For these parameters, Klemens-like recombination channels were always assumed ($\omega_i \approx \omega_0/2$). Errors are indicated below respective values. All quantities are given in cm$^{-1}$.}
	\begin{tabular}{cccccccc}
		
		\toprule
		Mode & \hspace{0.2cm} $\omega_0$ \hspace{0.2cm} & \hspace{0.2cm} $\omega_i$ \hspace{0.2cm} & \hspace{0.2cm}$\omega_j$\hspace{0.2cm}& \hspace{0.2cm}$\Delta_3$\hspace{0.2cm} & \hspace{0.2cm}$\Delta_4$\hspace{0.2cm} &\hspace{0.2cm} $\Gamma_3$\hspace{0.2cm} & \hspace{0.2cm}$\Gamma_4$ \hspace{0.2cm}\\
		\midrule
		TO$_\Gamma$     & \hphantom{$\pm$}133.3 & 67 & 45 & \hphantom{$\pm$}-1.26 	& \hphantom{$\pm$}-0.013& \hphantom{$\pm$}0.36 & \hphantom{$\pm$}0.17\\
		&\hphantom{11}$\pm$0.1&&&\hphantom{-}$\pm$0.1\hphantom{1}&\hphantom{-}$\pm$0.01\hphantom{1}&$\pm$0.06&$\pm$0.01\\
		LO$_\Gamma$     & \hphantom{$\pm$}150.2 & 75 & 50 & \hphantom{$\pm$}-1.1\hphantom{0}	& \hphantom{$\pm$}0\hphantom{.00}     &  \hphantom{$\pm$}0.9\hphantom{0} & \hphantom{$\pm$}0.18\\
		&\hphantom{15}$\pm$0.2&&&\hphantom{-}$\pm$0.2\hphantom{0}&$\pm$0.02&$\pm$0.15&$\pm$0.02\\
		%LO$_\Gamma$     & 150.2$\pm$0.2 & 75 & 50 & -1.09$\pm$0.18 	& 0$\pm$0.02    & 0.88$\pm$0.15 & 0.18$\pm$0.02\\
		\bottomrule
	\end{tabular}
\end{table}	
This model fits nicely our experimental data for the Raman shifts as well the experimentally determined broadening. The general behavior is quite similar for the TO and LO mode.\par
For both modes, the Raman shifts are dominated by the 3-phonon process, including a weaker lattice contribution. This lattice contribution matches quite exactly the PBEsol calculations for the LO mode, while for the TO mode a deviation is observed. Nevertheless, the trend is well resembled. A weak contribution of the 4-phonon process is found for the TO mode only.\par
The fit parameters describing the Raman shift and broadening of the TO and LO mode in Tab.~\ref{tab:phonon_decay_fit_params_O} are in agreement with the earlier results from Ref.~\cite{Serrano.2002}(under consideration of isotope effect and neglected lattice expansion) and as well with Ref.~\cite{He.2019} (considering the investigated temperature range). A detailed discussion is given in the supplementary.\par
The PBEsol results for the mode shift as a function of temperature of the TO and LO modes are much lower than the experimentally observed shift of the phonon modes. However, their contribution is approximately equal to the lattice contribution, which was calculated by the individual Grüneisen parameters (from Ref.~\cite{Serrano.2002}, derived via linear response method within DFT) and the experimental coefficient of thermal expansion (from Ref.~\cite{Nakamura.2022}). We use a quasi-harmonic model which only accounts for volume expansion to compute the frequency shift. Therefore, it is not surprising at all, that we only capture this effect. Hence, the PBEsol results can be associated with the lattice expansion contribution.\par 
Overall the analysis with the Klemens model~\cite{Klemens.1966} showed, that the 3- and 4-phonon processes and the lattice expansion contribution are relevant for the renormalization and the broadening of the respective phonon modes. This is in accordance with other publications~\cite{Serrano.2002,He.2019}. The lattice expansion contribution is quite well resembled by the quasi-harmonic model using PBEsol. This fact will be used for the identification and validation of harmonic and anharmonic contributions observed in the renormalization of acoustic phonons.
\subsection{Acoustic phonons}
\label{sec:acoustic_phonons}
\begin{figure*}  \centering
	\includegraphics[width=0.98\textwidth]{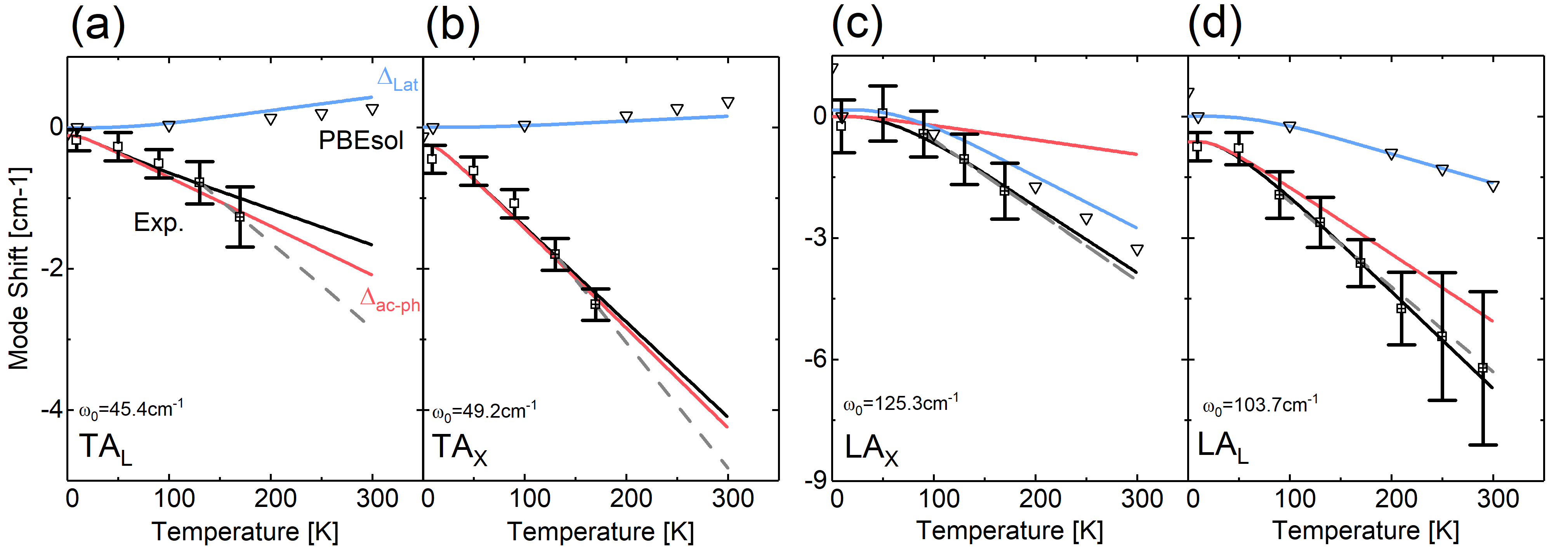}
	\caption{\label{fig:TA_LA_X_L_Center} Temperature dependent Raman shift of the second-order Raman scattering derived TA and LA phonons at the L- and X-point. The model for mode renormalization of optical phonons (black solid line, Eq.~(\ref{eq:center_fit_eq})) was applied to fit the data points by only considering a lattice expansion contributions (blue solid lines) and a acoustic-phonon-like process (red solid lines). PBEsol predictions of mode shifts are as well indicated (black open triangles). The data points used for linear extrapolation (grey dashed lines) are indicated (black crossed-squares).}
\end{figure*}

The experimentally determined TA$_\textup{L}$, TA$_\textup{X}$, LA$_\textup{L}$ and LA$_\textup{X}$ phonon energies at the respective temperature are shown in Fig.~\ref{fig:TA_LA_X_L_Center}(a-d). The experimental data for the LA$_\textup{L}$ mode shows a linear dependence, in agreement with results for other materials like Ge and NaI from Refs.~\cite{Kulda.2004,Li.2019,Kempa.2013}. The shift of the acoustic phonons from 10\,K up to RT is in the range of $2-6$\,\WNU{} and is significant smaller compared to the shifts of the optical ones.\par
The model by Klemens is only valid for zone-center optical phonons~\cite{Klemens.1966}. The observed shifts of the acoustic TA and LA phonons at the L and X point look in principle similar to the ones observed for the TO and LO phonons. Hence, we adapt this for a phenomenological description by including the known lattice expansion and Grüneisen parameters for the respective phonons at the critical points~\cite{Serrano.2002} ($\gamma($TA$_\textup{L}) = -0.9$, $\gamma($TA$_\textup{X}) = -0.3$, $\gamma($LA$_\textup{L}) = 1.5$ and $\gamma($LA$_\textup{X}) = 2.2$). An influence due to the choice of Grüneisen parameters can be excluded, because the deviations compared to our derived Grüneisen parameters is negligible (see supplementary). In this way, the chosen model is similar to Eq.~(\ref{eq:center_fit_eq}), while considering only the 3-phonon process (decay into two phonons), i.e. $\Delta_4=0$. This results in a linear dependence and was sufficient to describe the observed experiment data. To account for other linear processes in $\Delta_3$, we rename it to $\Delta_\textup{ac-ph}$. This now represents all anharmonic acoustic-phonon contributions. The lattice expansion contribution is included with the respective Grüneisen parameter. The $\omega_i$ values are chosen, as before for the optical phonons, as approximately the half of the respective acoustic phonon energy.\par
The model is shown as well in Fig.~\ref{fig:TA_LA_X_L_Center}(a-d) as solid line. The respective parts of lattice expansion (blue) and the contributions from the acoustic-phonon process (red) are shown. The PBEsol prediction of the mode shift is also indicated (black triangles), which coincides well with the predicted lattice contribution, as before for the optical modes. A linear extrapolation of the experimental data (dashed grey lines) shows negligible deviations compared to the model, especially for the LA phonons\,(i.e. $\lesssim$1cm$^{-1}$). This validates the extrapolated RT values for phonon energies of the \TAL{}, \TAX{} and \LAX{} modes. An extrapolation of the \LAX{} mode data points at higher temperatures can be done using the B$_1$ feature and the linear extrapolation of the $A_3$ feature. This agrees with the linear model up to RT. The unperturbed phonon energies are indicated by $\omega_0$ and listed in Tab.~\ref{tab:phonon_decay_fit_params_A}.\par
The validity of the model is indicated by a linear extrapolation, as shown in Fig.~\ref{fig:TA_LA_X_L_Center} (dashed grey line). Compared to the model fit, the linear extrapolation is almost identical for the LA phonons, while slight deviations are observed for the TA phonons (deviation at RT $\lesssim$1cm$^{-1}$).\par
\begin{table}
	\caption{\label{tab:phonon_decay_fit_params_A} Fit parameters (Eq.~(\ref{eq:center_fit_eq})) of acoustic phonons energetic temperature dependence. $\omega_0$ is the unperturbed phonon energy, $\omega_i$ the phonon energy of 3-phonon process like contribution and $\Delta_3$ the zero-temperature renormalization constant.}
	\begin{tabular}{cccc}
		\toprule
		Mode & \hspace{0.2cm} $\omega_0$ \hspace{0.2cm} & \hspace{0.2cm} $\omega_i$ \hspace{0.2cm} & \hspace{0.2cm}$\Delta_\textup{ac-ph}$\hspace{0.2cm} \\
		\midrule
		TA$_\textup{L}$ & 45.4  & 23   & -0.11   \\
		TA$_\textup{X}$ & 49.2  & 25   & -0.25   \\
		LA$_\textup{L}$ & 103.7 & 52   & -0.63   \\
		LA$_\textup{X}$ & 125.3 & 62   & -0.16   \\
		\bottomrule
	\end{tabular}
\end{table}
The model describes the experimentally observed behaviour quite well. Only the TA modes show slight deviations at low temperatures. This could be linked to possible 3-phonon decay channels, which exist for the LA phonon (LA$\rightarrow$TA+TA) but not for TA phonons~\cite{Okubo.1983}. For the TA phonons, only elastic interactions with thermal phonons can be expected~\cite{Kulda.2004}. The lattice contribution is negative for the TA mode, as observed for other zincblende semiconductors and is a consequence of the negative Grüneisen parameter~\cite{Serrano.2002,Gopakumar.2017}. Its effect is quite small compared to the acoustic-phonon contribution and negligible for the TA$_\textup{X}$ mode. The mode shifts for the TA and LA modes at both critical points are comparable in magnitude. The lattice contribution plays a considerable role for the LA modes, whereby it dominates for the LA$_\textup{X}$ mode.\par 
The relative deviations by the calculated lattice contribution and the quasi-harmonic contribution from the PBEsol calculations are significant smaller for the longitudinal phonons than for the transversal phonons. The same trend is observed for the optical phonons and may be related to the transversal characteristics of the respective phonon modes. However, the absolute deviations are negligible and hence the lattice contribution is quite well resembled by the PBEsol theory.\par
Under consideration of the uncertainties, the extrapolated and observed Raman shifts of the acoustic phonon modes at RT are in agreement with the results of the neutron scattering measurements from Ref.~\cite{Hennion.1972}, except for the TA$_\textup{L}$ mode. We predict a value of 43\,cm$^{-1}$ at RT, which is significant larger than 34\,cm$^{-1}$~\cite{Hennion.1972}(neutron scattering) or 32\,cm$^{-1}$~\cite{Brafman.1976}(linear extrapolation from pressure dependence of the hombohedral phase). For the neutron scattering measurements, this deviation is slightly larger than the given uncertainties. We also note the significant overestimation of the LO phonon energy from neutron scattering compared to Raman spectroscopy measurements~\cite{Potts.1973}. This may explain the large deviation observed for the TA$_\textup{L}$ mode.\par
Our determined TA mode positions and shifts can be supported by two different facts. Firstly, by comparing the linear extrapolation with inelastic neutron scattering measurements of CuI~\cite{Gopakumar.2017} (see supplementary). In this reference, a peak structure is observed at about 40\,cm$^{-1}$, which we associate with the TA modes at L- and X-point. Secondly, by considering the temperature dependence of the phonons for the halide group semiconductor CuBr. The relative shifts of the TA phonons for CuBr are matching quite well to the observed relative shifts of the TA phonons of CuI~\cite{Hoshino.1976}.\par% Fig.~\ref{fig_sup:TA_Own_and_INS_Data}
Our determined LA mode positions and shifts agree well for the L-point, while for the X-point some deviations are observed. The \mbox{[LO-LA]$_\textup{L}$} difference mode observed at 43\WNU{}(Ref.~\cite{Livescu.1986}) as well as the relative shifts for the CuBr~\cite{Hoshino.1976} support the \LAL{} assignment. Our observed relative shift for the \LAX{} mode is only 4\% and hence lower than the relative shift of CuBr's \LAX{} mode with 14\%~\cite{Hoshino.1976}. The strong background signal at the B structure may reason an underestimation of the \LAX{} mode shift.\par
\begin{figure}  \centering
	%C:\Daten\_Project 3 CuI\Data Analysis\Raman\220331 Construct 2DOS Spectrum\_6 Phonon Dispersion Compare.opj -->  Seifert Phonon Disp
	\includegraphics[width=0.49\textwidth]{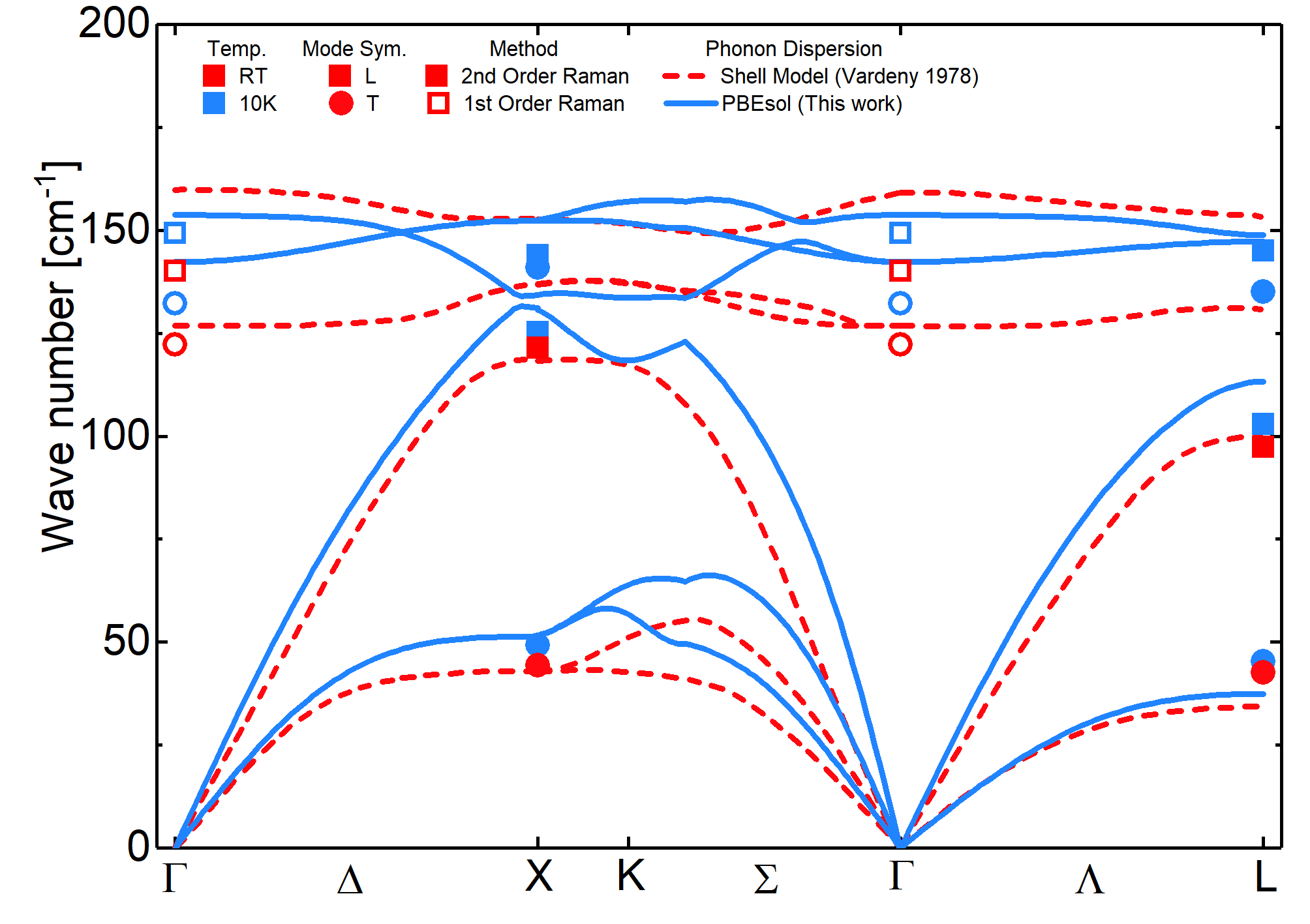}
	\caption{\label{fig:PhDisp_Overview_RT_10K} Phonon dispersion along high symmetry directions from the shell model~\cite{Vardeny.1977} (red dashed line, $T=$\,RT) and our PBEsol results (blue line, $T=10$\,K). Our experimental data from first- and second-order Raman scattering (open and full symbols). Red colors indicates RT and blue colors indicate a temperature of 10\,K. Longitudinal modes are indicated by squares and transversal ones with circles. Note the known significant deviations between Raman data and neutron scattering data (red dashed lines) for the TA$_\textup{L}$ and LO$_\Gamma$ modes~\cite{Brafman.1976,Prevot.1974,Hennion.1971}.}
\end{figure}

These results of our determined optic and acoustic phonon frequencies are summed up in Fig.~\ref{fig:PhDisp_Overview_RT_10K} and validate the observed temperature trend of both TA phonons and the \LAL{} phonons with measurements down to about 10\,K. Especially at elevated temperatures the linear trend is observed for all acoustic phonons up to RT. A model including lattice expansion and acoustic phonon processes could be used to describe this behavior. This seems to be valid at least up to RT and possibly as well above~\cite{Gopakumar.2017}. This is also consistent with CuI's high predicted and observed anharmonicity~\cite{Knoop.2023,Serrano.2002,He.2019}, which manifests as well in the characteristics of the acoustic phonons.\par

\section{Summary}
\label{sec:conclusion}
We analyzed the anharmonic contributions for acoustic phonons by second-order Raman measurements and DFT-based calculations. This was done exemplarily on bulk single crystals for the p-type semiconductor CuI. Via Raman measurements the fundamental TO and LO modes and 19 second-order Raman features were observed. By means of the DFT-based calculated 2-phonon density of states, the measured spectral shape was well reproduced and each second-order feature was assigned to a two-phonon mode. This allowed us to derive the energy of acoustic phonons in addition to their energetic temperature dependence.\par
In this way, the CuI's acoustic phonon energies were determined down to 10\,K by using a Raman setup, with much higher accuracy than the up to now available neutron scattering results. The known systematic deviation in these neutron scattering measurements, do not extend towards the determined acoustic phonon energies. Only for the \TAL{} phonon mode we determined a significant different energy.\par
The TO and LO phonons were analyzed by a Klemens model including harmonic lattice contributions. The latter were additionally verified by using DFT computations and extended for the acoustic phonons to determine their anharmonic contributions. Here the difference between experimental observed mode shift and predicted harmonic shift was assigned to the anharmonic contributions of the respective acoustic phonons. These anharmonicities dominate for most acoustic phonons, similar to the zone-center optical phonons. This trend is mostly linear and extends up to at least room temperature. Hence, a 3-phonon decay or the elastic thermal phonon scattering may be considered as relevant phonon-phonon interaction processes.\par
These findings are in agreement with earlier results and extend CuI's determined acoustic phonon properties to lower temperatures with higher accuracy. This approach as well shows that it is possible to characterize the acoustic anharmonicities via Raman scattering.
\vfil
\section{Acknowledgment}
%ACKNOWLEDGMENT
We thank Evgeny Krüger for discussion and Lukas Trefflich as well as Gabriele Benndorf (both from the University of Leipzig) for technical support. We gratefully acknowledge funding from the Deutsche Forschungsgemeinschaft (DFG, German Research Foundation) through FOR 2857 - 403159832. Furthermore we acknowledge the Leibniz Supercomputing Center for granting computational resources (Project No. pn68le). R.H. acknowledges the Leipzig School of Natural Sciences BuildMoNa.

%\section{References}
\bibliography{SecondOrderRamanCuI_Paper_Bibfile}

\end{document}

%\preprint{AIP/123-QED}

\title[Supp. Mat.: Det. of ac. phonon anharm. via second-order Raman scattering in CuI]{Supplemental Material: Determination of acoustic phonon anharmonicities via second-order Raman scattering in CuI}
% Force line breaks with \\

%\author{R. Hildebrandt $^1$, M. Seifert$^2$, Janine George$^{2,3}$, S. Blaurock$^4$, S. Botti$^{2,5}$, H. Krautscheid$^4$, M. Grundmann$^1$, C. Sturm$^1$}
%
%\address{1 Universität Leipzig, Felix Bloch Institute for Solid State Physics, Semiconductor Physics Group, Linn\'{e}stra{\ss}e 5, 04103 Leipzig }
%\address{2 Friedrich-Schiller-Universität Jena, Institute of Condensed Matter Theory and Optics, Max-Wien-Platz 1, 07743 Jena}%
%\address{3	Federal Institute for Materials Research and Testing 	Department Materials Chemistry, Unter den Eichen 87, 12205 Berlin}%
%\address{4	Univerität Leipzig, Institute for Inorganic Chemistry, Johannisallee 29, 04103 Leipzig}
%\address{5	Ruhr University Bochum, Research Center Future Energy Materials and Systems of the Research Alliance Ruhr, Faculty of Physics and ICAM, Universitätstra{\ss}e 150, 44780 Bochum}%

\author{R. Hildebrandt} \email{ron.hildebrandt@uni-leipzig.de}
%\altaffiliation[Also at ]{Bottom page footnote}%Lines break automatically or can be forced with \\
\affiliation{ 
	Universität Leipzig, Felix Bloch Institute for Solid State Physics, Semiconductor Physics Group, Linn\'{e}stra{\ss}e 5, 04103 Leipzig
}

\author{M. Seifert}
\affiliation{%
	Friedrich-Schiller-Universität Jena, Institute of Condensed Matter Theory and Optics, Max-Wien-Platz 1, 07743 Jena
}%

\author{J. George}%Janine George
\affiliation{%
	Federal Institute for Materials Research and Testing
	Department Materials Chemistry, Unter den Eichen 87, 12205 Berlin}%
\affiliation{%
	Friedrich-Schiller-Universität Jena, Institute of Condensed Matter Theory and Optics, Max-Wien-Platz 1, 07743 Jena
}%

\author{S. Blaurock}
\affiliation{%
	Univerität Leipzig, Institute for Inorganic Chemistry, Johannisallee 29, 04103 Leipzig
}

\author{S. Botti}
\affiliation{%
	Friedrich-Schiller-Universität Jena, Institute of Condensed Matter Theory and Optics, Max-Wien-Platz 1, 07743 Jena
}%
\affiliation{%
	Ruhr University Bochum, Research Center Future Energy Materials and Systems of the Research Alliance Ruhr, Faculty of Physics and ICAM, Universitätstra{\ss}e 150, 44780 Bochum
}%

\author{H. Krautscheid}
\affiliation{%
	Univerität Leipzig, Institute for Inorganic Chemistry, Johannisallee 29, 04103 Leipzig
}%
%
%\author{A. Dadgar}
%\affiliation{%
	%	Otto-von-Guericke-Universität Magdeburg, Institute for Physics, Abteilung Halbleiterepitaxie, Universitätsplatz 2, 39106 Magdeburg 
	%}%

\author{M. Grundmann}
%\homepage{http://www.Second.institution.edu/~Charlie.Author.}
\affiliation{ 
	Universität Leipzig, Felix Bloch Institute for Solid State Physics, Semiconductor Physics Group, Linn\'{e}stra{\ss}e 5, 04103 Leipzig
}

\author{C. Sturm}%
\affiliation{ 
	Universität Leipzig, Felix Bloch Institute for Solid State Physics, Semiconductor Physics Group, Linn\'{e}stra{\ss}e 5, 04103 Leipzig
}

\date{\today}% It is always \today, today,
%  but any date may be explicitly specified

\maketitle

%\section{Supplementary}

\renewcommand{\thefigure}{S\arabic{figure}}
\setcounter{figure}{0}
\section{Difference modes spectrum}
\begin{figure}[htp] \centering
	\includegraphics[width=0.49\textwidth]{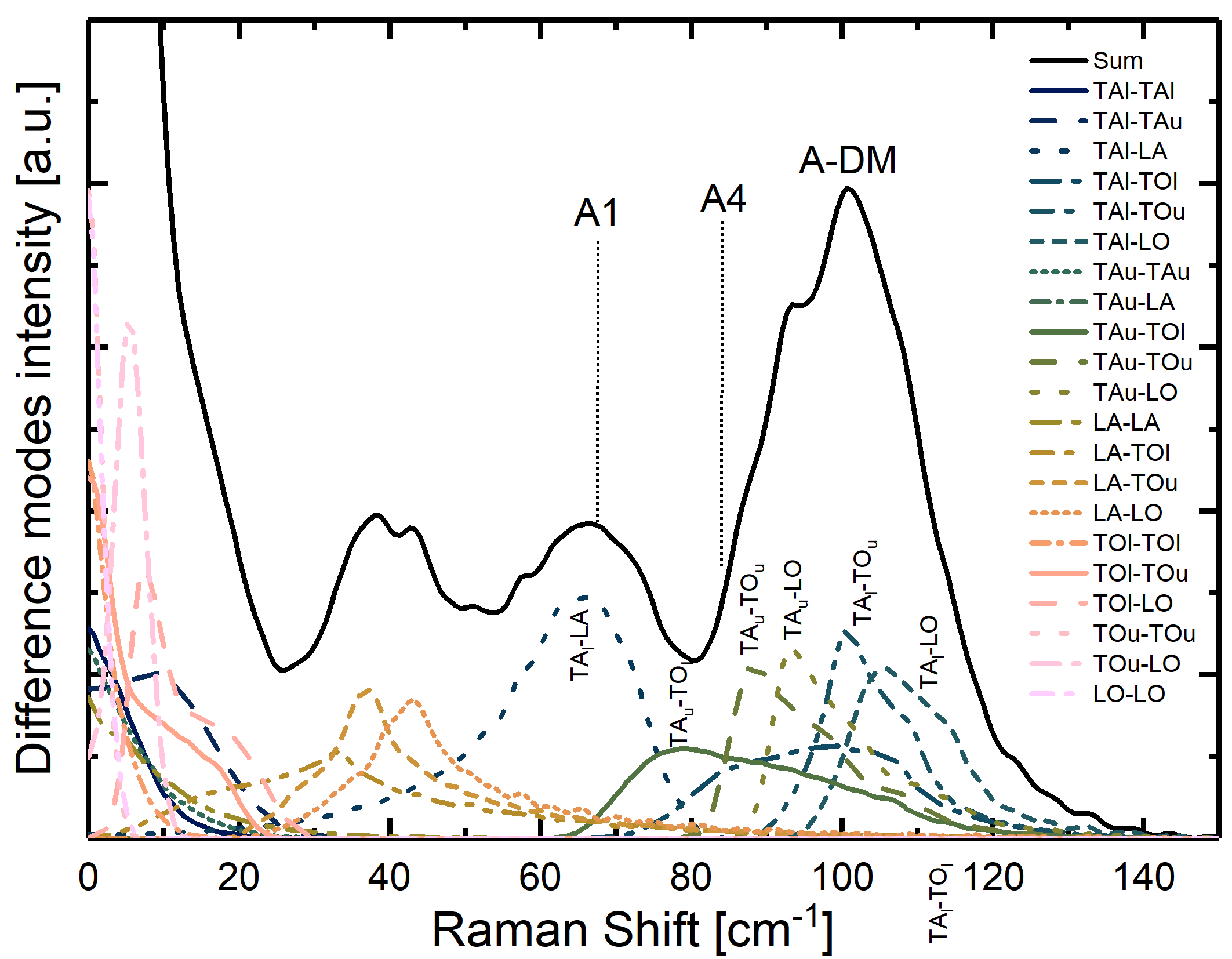}
	\caption{\label{fig_sup:diff_modes} Calculated 2-Phonon density of states for difference modes. The assigned difference mode features are indicated by vertical dotted lines and labels, as well as the group of optical and transversal acoustic difference modes, labeled A-DM. Due to the strong overlap, the transversal modes were distinguished into upper (u) and lower (l) modes to account the degeneracy lifting along the $\Sigma$-direction.}
\end{figure}

The manuscript covers mostly the overtone and combination 2-Phonon modes. For completeness and support of the assignment of features A$_1$ and A$_4$, the 2-Phonon density of states difference modes spectrum is displayed in Fig.~\ref{fig_sup:diff_modes}. The group of optical and transversal acoustic difference modes, labeled as A-DM, is observable at elevated temperatures and is negligible at 10\,K. At 10\,K the transversal acoustic overtones dominate the spectrum.

\section{Optical phonons}
\label{sec:optical_phonons_old}
\begin{figure}[htp] \centering
	%location: C:\Daten\_Project 3 CuI\Data Analysis\Raman\220331 Construct 2DOS Spectrum\_2 Check_Determined_2Phonon_features.opj --> 13 Overview Fit Center TA LA O for Paper
	\includegraphics[width=0.49\textwidth]{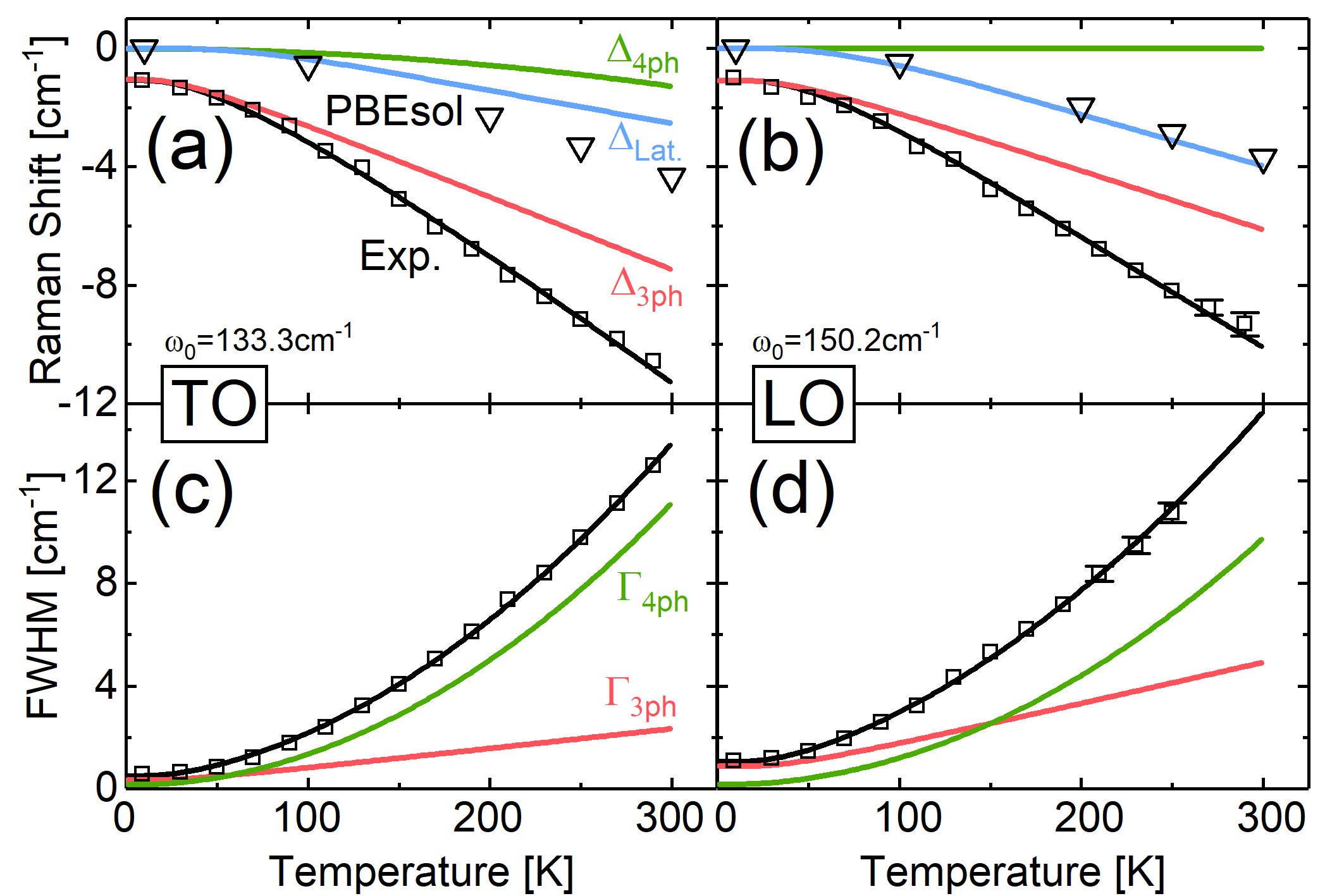}
	\caption{\label{fig:TO_LO_Center_and_width} Experimental Raman shift for the TO (a) and LO (b) mode as well as the FWHM for the TO (c) and LO (d) mode in dependence of the temperature (open squares). A model fit (black) with the individual contributions of the 3-phonon process $\Delta_3$ (red), 4-phonon process $\Delta_4$ (green) and the lattice $\Delta_\textup{Lat.}$ (light blue) describes the observations. The quasi-harmonic Raman shifts are predicted by PBEsol (black open triangles). $\omega_0$ is the unperturbed phonon energy. Error bars not indicated are smaller than the symbol size.}
\end{figure}

The energy and broadening of the TO and LO mode is obtained by means of Lorentzian peak functions as a function of temperature shown in Fig.~\ref{fig:TO_LO_Center_and_width}.\par
This data was described by Eq.~2 of the main text for the frequency shift and by Eq.~(\ref{eq:width_fit_eq}) for the broadening of the respective phonon modes. Anharmonic broadening due to phonon-phonon interaction is given by:
\begin{eqnarray}
	\label{eq:width_fit_eq}
		\Gamma(T) =& \, \Gamma_{3\textup{ph}}(T) + \Gamma_{4\textup{ph}}(T),\\ 
		\textup{with}&\nonumber\\
		\Gamma_{3\textup{ph}}(T) =& \ \Gamma_3\cdot[1+\sum_{i=1}^{2}\nolimits n(\omega_i)] \nonumber \\
		\Gamma_{4\textup{ph}}(T) =& \ \Gamma_4\cdot[1+\sum_{j=1}^{3}\nolimits n(\omega_j) + n^2(\omega_j)] \nonumber
\end{eqnarray}
Here $\Gamma_3$ and $\Gamma_4$ are the zero-temperature anharmonic broadening constants, which are related to their type of contribution respectively.\par
The model (black lines) describes these changes of the optical phonons frequency shift and width quite well. The individual contributions from the lattice expansion (light blue), 3-phonon (red) and 4-phonon (green) processes are shown as solid lines. The quasi-harmonic Raman shift predicted by PBEsol (black open triangles) calculations are indicated.\par
We now shortly discuss the model fit parameters in context of results from other reports by starting with the mode shift and continuing then with the broadening. The ratio $\Delta_{3,\textup{TO}}$ / $\Delta_{3,\textup{LO}}$ $\approx$ 1 is similar to the one reported in Ref.~\cite{Serrano.2002}. Our result of $\Delta_{4}$ $\approx$ 0 supports their approach to neglect 4-phonon processes. The difference in absolute values $\Delta_3$, i.e. our result of $1.3$\,\WNU{} and $1.1$\,\WNU{} and their result of $\approx2.65$\,\WNU{} and $\approx2.3$\,\WNU{}, can be explained by the fact that a contribution due to the lattice expansion, was not included in Ref.~\cite{Serrano.2002}. This reasons their larger values, which compensates the lattice expansion contribution.\par
The results does not significantly depend on the choice of the $\omega_i$ or $\omega_j$ parameters. Hence one can not exclude the presence of multiple decay channels, which have comparable $\omega_i$ values. For example LO(150\,\WNU{}) $\rightarrow$ LA(100\,\WNU{}) + TA(50\,\WNU{}) is expected to be present in similar magnitude as LO(150\,\WNU{}) $\rightarrow$ LA(75\,\WNU{}) + LA(75\,\WNU{}). Only for extreme cases a differentiation is possible. These would require the observation of a stronger pronounced kink in the Raman shift. In this way, a significant contribution of energetically strongly asymmetric decay channels can be excluded such as, for example, found for the LO(150\,\WNU{}) $\rightarrow$ TO(132\,\WNU{}) + TA(18\,\WNU{}) channel.\par
The broadening of the TO and LO modes behave as well quite similar. The 4-phonon process dominates the linewidth increase at large temperatures while, at low temperatures, the 3-phonon process is dominant. This is especially the case for the LO mode, which results in the larger zero-temperature linewidth of 1.03\,cm$^{-1}$ compared to 0.53\,cm$^{-1}$ for the TO mode. These values agree well with the ones reported in Ref.~\cite{Serrano.2002}. At elevated temperatures, the TO and LO width characteristics are reproduced~\cite{Serrano.2002,He.2019} (considering the copper isotope distribution).
\section{Thermal expansion coefficient}
\begin{figure}[htp] \centering
	\includegraphics[width=0.39\textwidth]{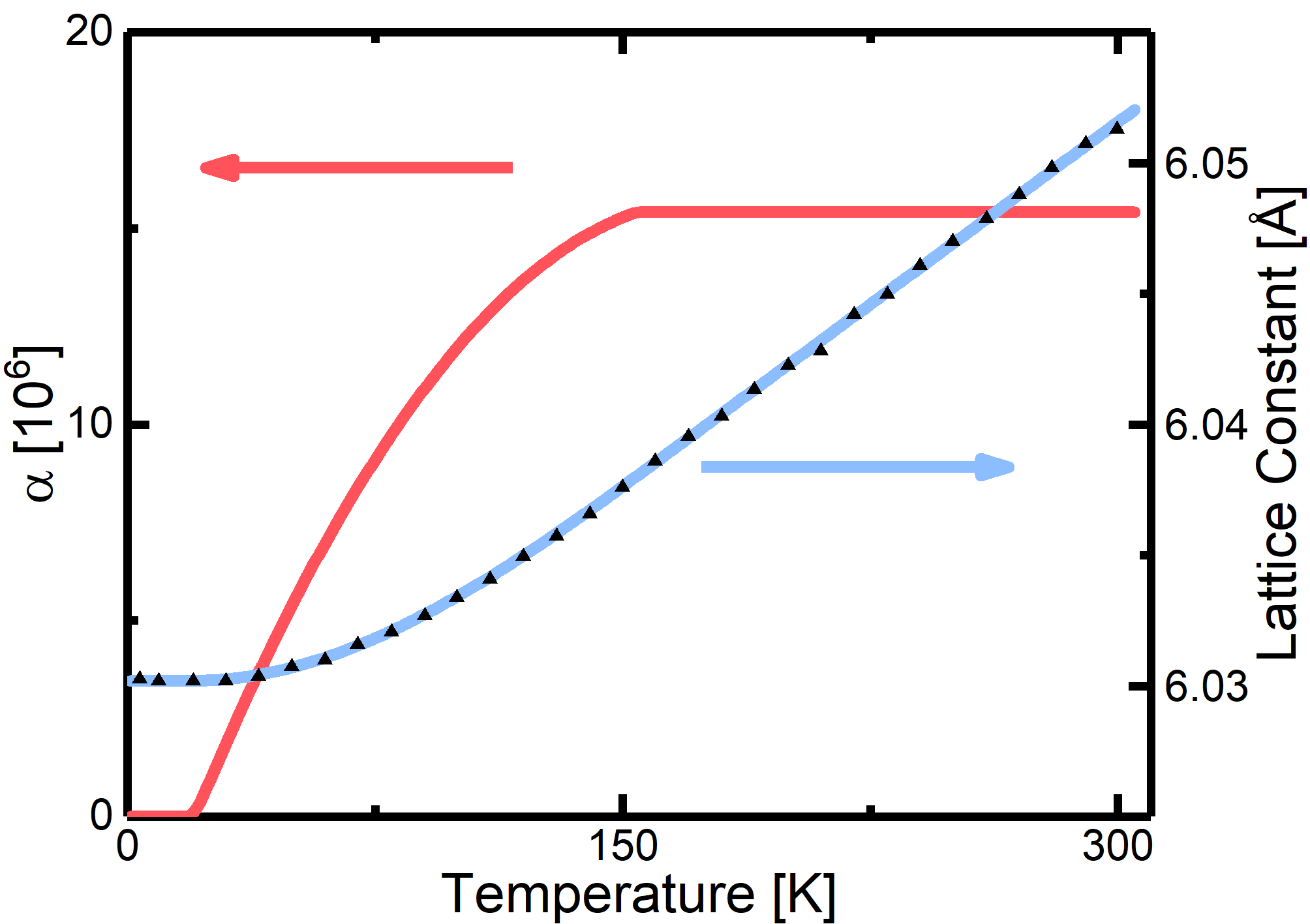}
	\caption{\label{fig_sup:alpha_nakamura_and_lattice_constant} The experimental data represents the lattice constants (triangles) and the data is taken from Ref.~\cite{Nakamura.2022}. The respective used lattice constant is shown (blue line), which yields the derived coefficient of thermal expansion (red line).}
\end{figure}

The shift of the phonon mode is determined by the Grüneisen parameters and the coefficient of thermal expansion (CTE). The used CTE is shown in Fig.~\ref{fig_sup:alpha_nakamura_and_lattice_constant} and was derived by using the bulk lattice constant measured in Ref.~\cite{Nakamura.2022}. This CTE is in accordance with other results~\cite{Gopakumar.2017, Plendl.1972}.\par

\section{TA phonon temperature dependence}
\begin{figure} \centering
	\includegraphics[width=0.49\textwidth]{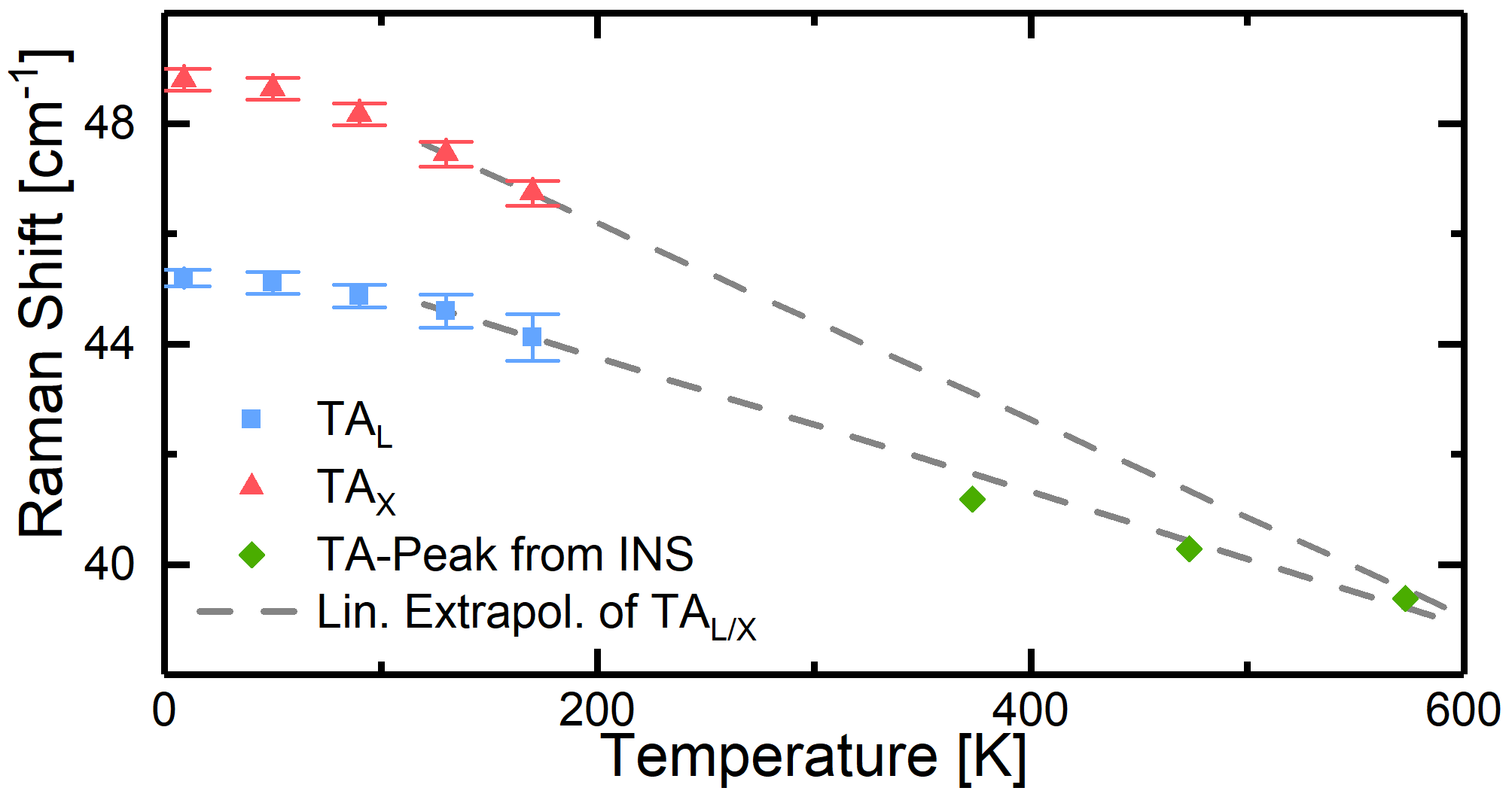}
	\caption{\label{fig_sup:TA_Own_and_INS_Data} Second-order Raman scattering determined phonon energies of both TA modes and the TA peak positions at elevated temperatures from the inelastic neutron scattering~\cite{Gopakumar.2017}. The linear extrapolation of our determined TA phonon energies agrees well with their TA peak positions.}
\end{figure}

The extrapolation of the TA modes to higher temperatures and comparison with results from neutron scattering from Ref.~\cite{Gopakumar.2017} is shown in Fig.~\ref{fig_sup:TA_Own_and_INS_Data}. The extrapolation supports the identified temperature dependence of TA phonons. 
%\clearpage

\section{Additional information about phonon calculation }

\begin{figure}[htp] \centering
	\includegraphics[width=0.49\textwidth]{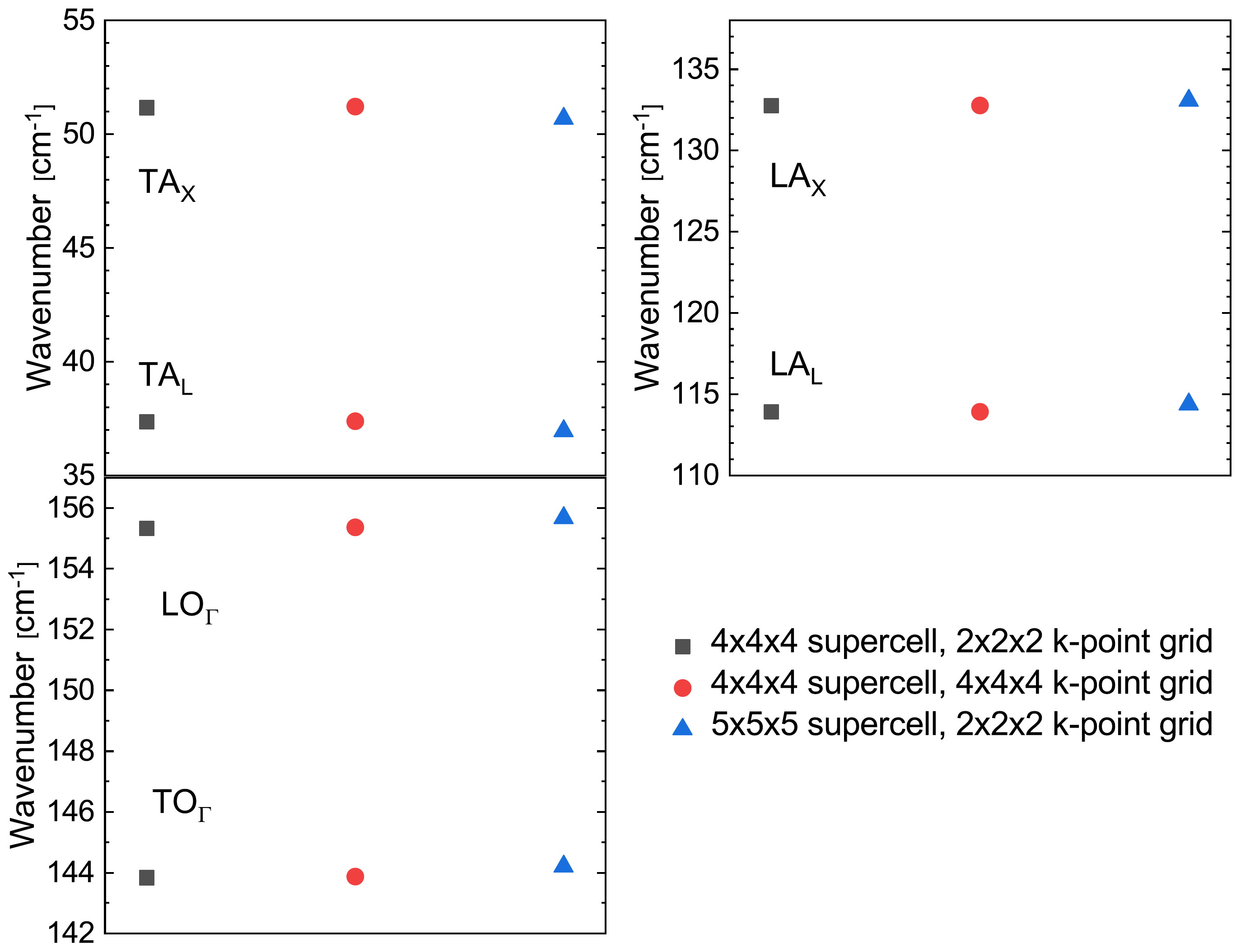}
	\caption{\label{fig_sup:PhononConvergence} Influence of the supercell size and the corresponding $\mathbf{k}$-point grid on the wavenumbers of exemplary phonon modes at certain high-symmetry point in the Brillouin zone. Calculated at $T$=0\,K.}
\end{figure}

Important quantities for the phonon calculations are the size of the supercell as well as the corresponding $\mathbf{k}$-point grid for the computation of the forces as part of the finite displacement method. To discuss the influence of those regarding the results shown in the manuscript the convergence of certain modes at certain high-symmetry points of the Brillouin zone are shown in Fig.~\ref{fig_sup:PhononConvergence}. These specific modes are chosen, since they were also discussed in the main manuscript. It is shown, that both increasing the supercell size as well as the density of the $\mathbf{k}$-point grid causes differences smaller than 1~\%. 
\begin{figure}[htp] \centering
	\includegraphics[width=0.49\textwidth]{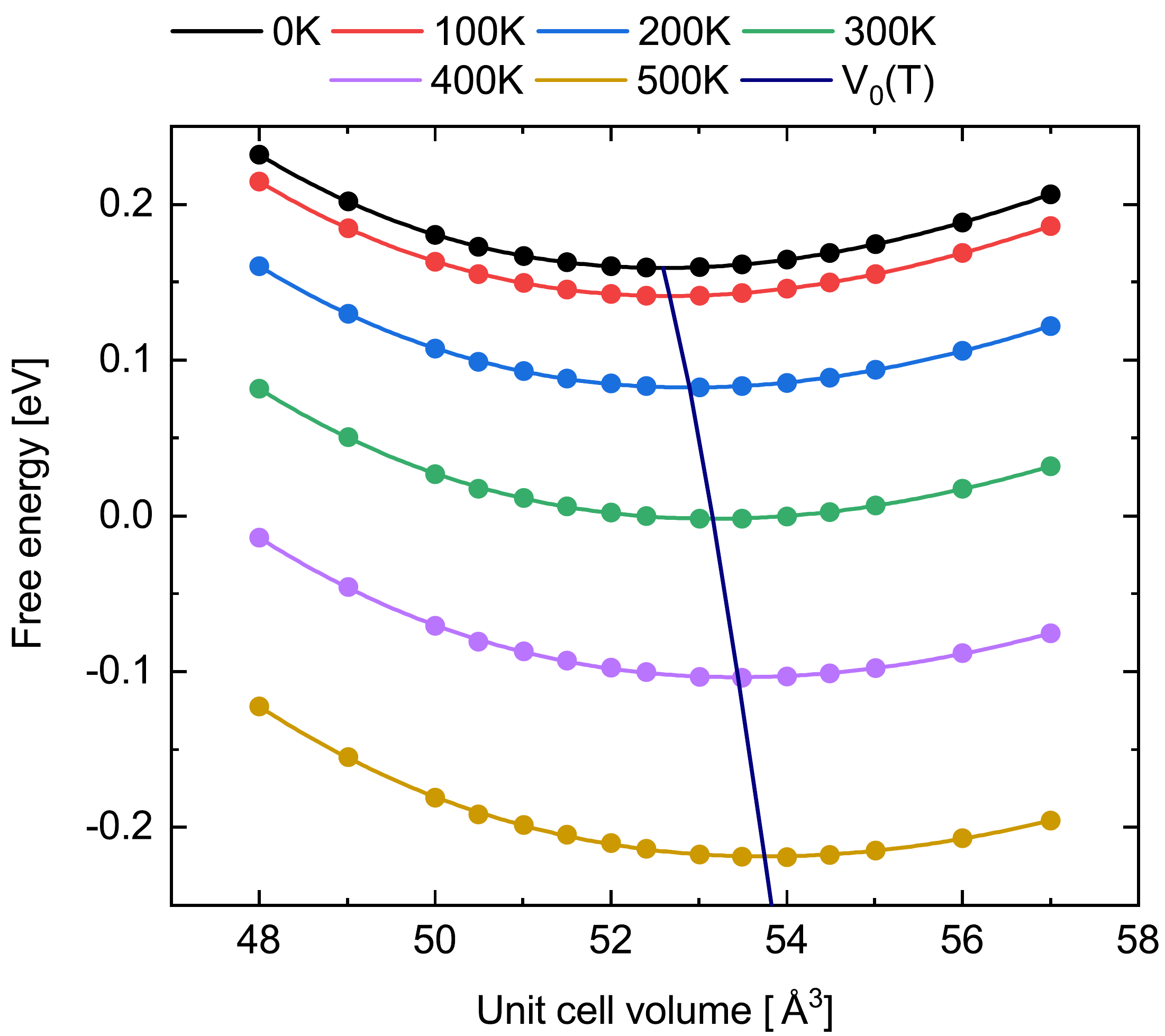}
	\caption{\label{fig_sup:EquationOfStatesFit} Determination of the temperature-dependent volume via fitting a Vinet equation of states. }
\end{figure}
To compute temperature-dependent phonon modes, we applied the quasi-harmonic approximation to obtain the free energy at different volumes. Then the temperature-dependent volume of the unit cell was obtained by fitting a Vinet equation of states to the corresponding temperature curves. These steps were performed using phonopy~\cite{Togo.2008, Togo.2015}. In Fig.~\ref{fig_sup:EquationOfStatesFit} those curves are exemplarily shown for some temperatures as well as the development of the equilibrium volume.

\section{Grüneisen Parameter comparison}
\begin{figure}[htp] \centering
	\includegraphics[width=0.49\textwidth]{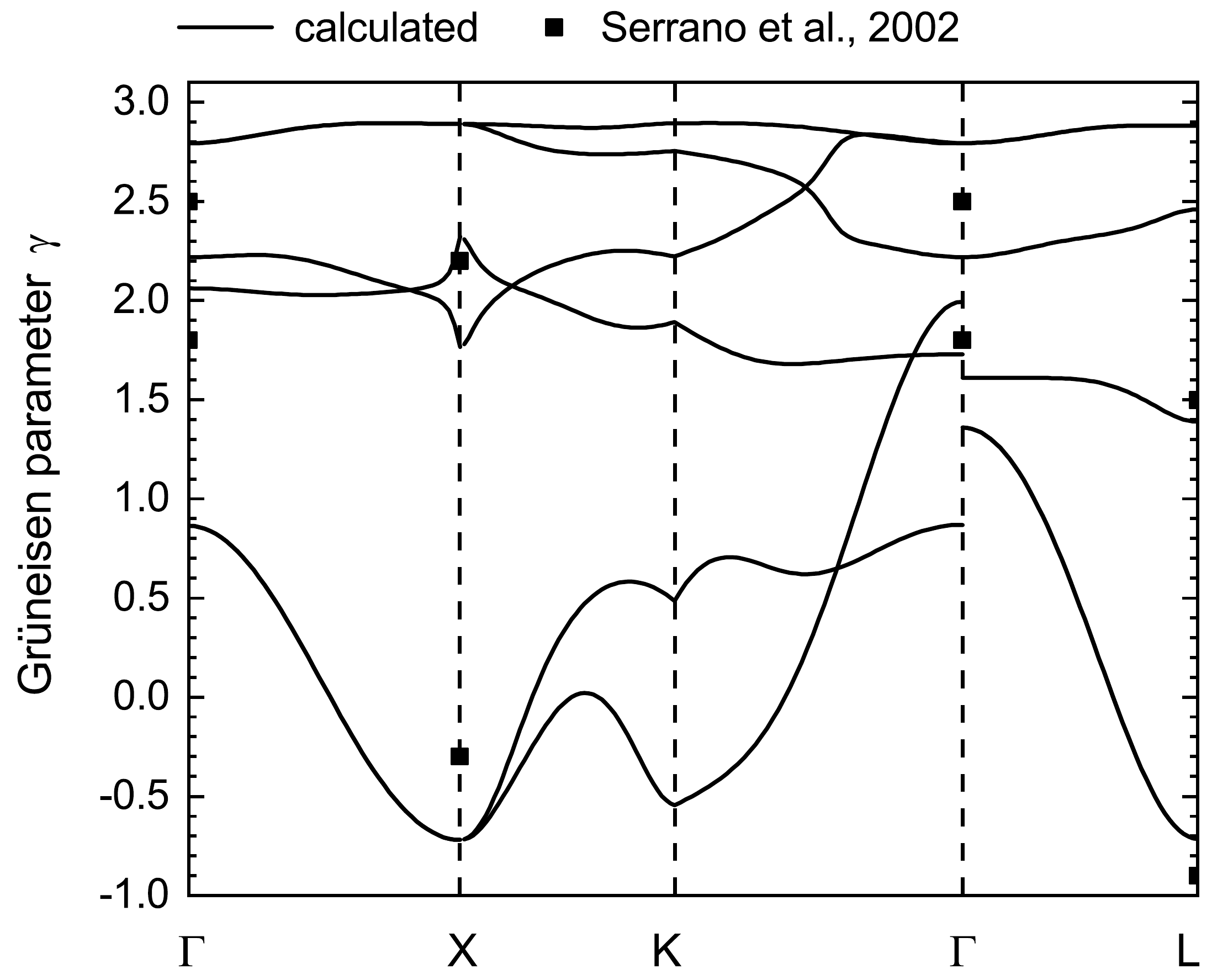}
	\caption{\label{fig_sup:GE-Parameter-BS} Our calculated Grüneisen band structure of CuI. Square symbols are Grüneisen parameters from Ref.~\cite{Serrano.2002} used to calculate $\Delta_{\textup{Lat.}}$: $\gamma($TO$_\Gamma) = 1.8$,  $\gamma($LO$_\Gamma) = 2.5$, $\gamma($TA$_\textup{L}) = -0.9$, $\gamma($TA$_\textup{X}) = -0.3$, $\gamma($LA$_\textup{L}) = 1.5$ and $\gamma($LA$_\textup{X}) = 2.2$.}
\end{figure}

In the manuscript, we mentioned negligible deviations of our calculated Grüneisen parameter $\gamma$ at specific high-symmetry points to  the literature values. This is displayed via the Grüneisen band structure of CuI in Fig.~\ref{fig_sup:GE-Parameter-BS} including the respective literature values.

\clearpage

%\section{References}
\bibliography{SecondOrderRamanCuI_Paper_Bibfile}
%\bibliography{SecondOrderRamanCuI_Paper_Bibfile.bib}

%\nocite{*}
%\bibliography{anisotropic_raman_references}% Produces the bibliography via BibTeX.